\documentclass[11pt]{article}

\usepackage[final]{acl}

\usepackage{times}
\usepackage{latexsym}
\usepackage[T1]{fontenc}
\usepackage[utf8]{inputenc}
\usepackage{microtype}
\usepackage{inconsolata}
\usepackage{graphicx}

\usepackage{url}
\usepackage{hyperref}
\usepackage{tikz}

\usepackage{pifont}
\usepackage[dvipsnames]{xcolor} % 方便用 ForestGreen / BrickRed 等命名色
\usepackage{booktabs}
\usepackage{array}
\usepackage{tabularx}

\usepackage{threeparttable}
\usepackage{multirow}

\usepackage{subcaption}

\usepackage{longtable}
\usepackage{caption}

\usepackage[dvipsnames]{xcolor}

\usepackage{amsmath}
\usepackage[nameinlink,noabbrev]{cleveref}

\usepackage{bbm}

\crefname{figure}{Figure}{Figures}
\Crefname{figure}{Figure}{Figures}

\crefname{table}{Table}{Tables}
\Crefname{table}{Table}{Tables}

\usepackage{wrapfig}
\usepackage{needspace}

\usepackage{empheq}
\usepackage{listings}
\usepackage[most]{tcolorbox}
\usepackage{adjustbox}

\usepackage{fvextra}

\newcommand{\eval}{\mathsf{eval}}
\newcommand{\complete}{\mathsf{comp}}
\newcommand{\correct}{\mathsf{corr}}

\newcommand{\projName}{\textsc{PostcondBench}}

\newcommand{\cmark}{\textcolor{ForestGreen}{\ding{51}}} % ✓
\newcommand{\xmark}{\textcolor{BrickRed}{\ding{55}}}    % ✗

\newcommand{\hmark}{%
  \leavevmode
  \ooalign{%
    \hfil\textcolor{Orange}{\ding{51}}\hfil\cr
    \hfil\raise0.15ex\hbox{\textcolor{Orange}{\scriptsize\ding{55}}}\hfil\cr
  }%
}

\newcommand{\camchg}[1]{{#1}}
\newenvironment{camchgblock}{\begingroup}{\endgroup}
\newcommand{\camcaption}[1]{\texorpdfstring{{#1}}{#1}}

\newcommand{\camchgtwo}[1]{{#1}}
\newenvironment{camchgblocktwo}{\begingroup}{\endgroup}
\newcommand{\camcaptiontwo}[1]{\texorpdfstring{{#1}}{#1}}

\newtcolorbox{takeawaybox}{
  enhanced,
  boxrule=0.6pt,
  colback=black!3,
  colframe=black,
  arc=2pt,
  left=6pt,right=6pt,top=6pt,bottom=6pt
}

\title{\projName: Benchmarking Correctness and Completeness in Formal Postcondition Inference}

\author{Gehao Zhang \and Juan Zhai \\
  Manning College of Information and Computer Sciences \\
  University of Massachusetts Amherst \\
  Amherst, MA, USA \\
  \texttt{\{gehaozhang,juanzhai\}@umass.edu}}

\begin{document}

\maketitle

\begin{abstract}

Formal postconditions precisely characterize program behavior and support debugging, testing, and verification, but writing them requires substantial expertise and effort.
This has motivated recent work on automatically generating postconditions from code and natural-language artifacts using large language models (LLMs).
However, evaluation remains a key bottleneck.
Existing benchmarks primarily emphasize correctness under limited evaluation settings, often relying on surface-form matching or manual assessment on small or synthetic datasets.
% While prior work primarily assesses \emph{correctness}—whether postconditions hold on correct executions—\emph{completeness}, i.e., whether postconditions capture a method’s key behaviors, is difficult to evaluate automatically.
% In practice, many semantically valid postconditions differ substantially in surface form, making reference-based matching brittle and pushing existing evaluations toward manual assessment or small synthetic benchmarks.

We introduce \projName, a multilingual benchmark for evaluating method-level postcondition generation from real-world software.
\projName~comprises 420 Python and Java tasks drawn from 121 open-source projects, each paired with a high-quality ground-truth postcondition set constructed with expert involvement.
To enable automatic evaluation, \projName~provides a runnable execution environment and operationalizes completeness via \emph{defect discrimination}: a postcondition set is more complete if it is violated by more defective implementations while remaining satisfied on correct executions.
Using \projName, we formulate three generation settings and evaluate five SOTA LLMs.
Our results reveal a substantial gap between correctness and completeness, and show that repository-level dependencies and method complexity exacerbate this gap.

\end{abstract}

% \section{\juan{Follow my suggestion given below:}}

% We have NL-code-spec. We support both nl to spec and code to spec, kind of like two different benchmarks.
% \subsection{sampling}

% \begin{itemize}
%     \item comprehensive natural language
%     \item non-trival code
%     \item different dependencies. we can have standalone methods, but have to provide methods with different depths of dependencies.
%     \item diversity: avoid exactly the same NL, or at least not too many. 
%     \item test coverage, line/branch/path(???)
% \end{itemize}

% \subsection{spec generation}
% Use both NL and code as input to prompt LLMs. We can use several different LLMs and combine their results. Use filter to remove duplicates, both syntax level and semantic level. 

% \subsection{evaluate spec}

% \subsubsection{correctness}
% Mutate specifications to see whether test cases are comprehensive enough to distinguish a spec from its mutants. If not able to kill spec mutants, manually add test cases.

% \subsubsection{vacuity}
% Cannot kill any code mutant.

% \subsubsection{completeness}
% Generate code mutants, using both traditional methods and LLMs. 

% \subsection{evaluation}

% RQ1: how well do LLMs perform on NL to spec?

% RQ2: how well do LLMs perform on code to spec?

% RQ3: the different factors that affect the performance, including dependency complexity, providing context or not, examples in prompt, etc.

% RQ4: different prompt techniques.

% \juan{=======}

\section{Introduction}

A formal specification expresses the properties that a program is expected to satisfy~\citep{lamsweerde2000formal}.
By precisely characterizing program behavior, specifications play a central role in debugging, testing, and verification.
Despite their importance, formal specifications are scarce in real-world software, as writing them is time-consuming, error-prone, and requires substantial expertise~\citep{lamsweerde2000formal,snook2001exploring,henkel2008developing}.
As a result, a long line of work has studied the automatic generation of specifications from source code and natural-language (NL) artifacts, aiming to reduce the cost and expertise required for writing specifications~\citep{goffi2016automatic,blasi2018translating,zhai2020c2s,zhang2020automated,xie2022docter,pandita2012inferring}.

Large language models (LLMs) have significantly improved generative capabilities for software engineering tasks, including code generation and testing~\citep{chen2021evaluating,islam2024mapcoder,xue2024llm4fin}.
Several studies have shown that LLMs can also generate method-level postconditions directly from code or NL descriptions~\citep{xie2025effective,endres2024can,ma2025specgen}.
However, \emph{how to reliably evaluate generated postconditions} remains an open challenge.
% —especially with respect to their \emph{completeness}, \emph{i.e.}, whether they capture a method’s key behaviors rather than checking only a small subset.
%

Most existing automatic evaluations focus on \emph{correctness}.
Under sufficient test inputs, a postcondition is deemed correct if it is never violated when executed against a correct implementation~\citep{zhai2020c2s,endres2024can}.
Correctness alone is insufficient as a correct postcondition set can still overlook important behavioral aspects of the method, i.e., they can be incomplete.
% Such postconditions are correct but incomplete.

Evaluating \emph{completeness} is fundamentally difficult because there is no single canonical specification.
The same behavioral property can often be expressed by many syntactically different but semantically equivalent postconditions~\citep{endres2024can,xie2025effective}.
This diversity makes matching against a fixed ground truth, e.g., via string or AST similarity, brittle and prone to penalizing semantically correct specifications.
Consequently, prior work has largely relied on manual assessment for evaluating completeness~\citep{xie2025effective,zhai2020c2s,ma2025specgen} (examples in \autoref{fig:sorting}).
Existing attempts at automatic completeness evaluation remain limited.
For instance, \citet{endres2024can} propose a completeness metric, but evaluate it only on HumanEval+~\citep{liu2023your}, a small dataset of standalone tiny Python programs.
To date, there is no benchmark that simultaneously provides
(i) multilingual, real-project-derived tasks,
(ii) high-quality ground-truth postconditions, and
(iii) automatic support for correctness and completeness evaluation.

We present \projName, a multilingual benchmark for evaluating the correctness and completeness of method-level postconditions, a widely used form of formal specification that constrains program states after method execution~\citep{lamsweerde2000formal}.
\projName~comprises 420 postcondition generation tasks in Python and Java, drawn from 121 diverse and popular open-source projects.
Each task is paired with a comprehensive, high-quality postcondition set constructed via a semi-automated pipeline with domain-expert involvement.
Moreover, \projName~provides a runnable execution environment with abundant context and tests, together with code-coverage analysis to ensure sufficiently exercised executions.

Crucially, \projName~operationalizes \emph{completeness} via \emph{defect discrimination}:
a postcondition set is more complete if it distinguishes more defective implementations from the reference set 
% by being violated on diverse defects 
while remaining satisfied on correct executions.
This formulation enables automatic computation 
% of a bug-completeness metric 
without relying on brittle equivalence checking against a single canonical specification.

Using \projName, we formulate three postcondition generation tasks and evaluate five state-of-the-art (SOTA) LLMs.
Our results show that completeness remains challenging even for strong models (e.g., at most 17\% in Python and 43\% in Java for GPT-5), and that there is a substantial gap between being \emph{correct} and being \emph{complete}.
We further analyze common causes of incorrect postconditions and the behavioral aspects most frequently overlooked by incomplete ones, highlighting concrete directions for future research.

Our main contributions are:
\begin{itemize}
    \item \projName, a multilingual benchmark of 420 Python/Java method-level postcondition generation tasks from 121 real-world repositories, with high-quality ground-truth postconditions.
    Unlike prior benchmarks based on standalone programs, \projName~captures repository-level context, including inter-method dependencies and non-trivial method complexity.
    We quantitatively characterize these factors via dependency analysis and method size (MLOC).
    \item An automatic evaluation platform that supports both correctness and completeness, with completeness measured via defect-discriminative power.
    \item Extensive evaluation of five SOTA LLMs on three generation tasks reveals a substantial gap between correctness and completeness, and shows that repository-level dependencies and method complexity further widen this gap.
\end{itemize}

% \begin{camchgblocktwo}
% \paragraph{Benchmark scope.}
% \projName~is designed to evaluate the quality of formal postconditions, while the reported scores can be a prerequisite capability for downstream uses such as verification and debugging.
% \projName~evaluates correctness and completeness by execution-based metrics with test cases and mutants.
% % We restrict the benchmark target methods to be well-covered by test cases due to the execution-based nature.
% % We also apply moderate data filtering during benchmark construction due to limitations in current specification frameworks or domain experts' judgment.
% \end{camchgblocktwo}

% \input{src/introduction}

\section{Background and Related Work}
\label{sec:background-related}

% \subsection{Formal Specifications and Postconditions}

% Formal specifications describe intended program behavior precisely and are
% widely used for verification, testing, and program understanding.
% Among them, \emph{postconditions} specify constraints on outputs and post-state
% properties after method execution, forming the basis of contract-based
% frameworks such as JML and icontract.
% Unlike tests, postconditions aim to characterize behavior independently of
% specific inputs.
% However, they admit many semantically valid formulations with varying strength,
% making both generation and evaluation inherently challenging.

% \subsection{Specification Generation}

Prior work on specification generation includes static analysis, dynamic analysis,
and data-driven approaches.
Static techniques infer specifications from program structure,
often providing soundness guarantees but struggling with scalability and
precision~\citep{Chen2016SupportingOC,Shoham2007StaticSM,Flanagan2001HoudiniAA}.
Dynamic approaches learn likely specifications from execution traces, but their
effectiveness depends heavily on test coverage~\citep{ernst2001dynamically,nimmer2002automatic}.
Repository-mining methods exploit recurring patterns across large codebases, yet
are sensitive to repository quality and consistency~\citep{ramanathan2007static,nguyen2014mining}.

Recently, LLMs have been applied to generate formal
specifications from NL and/or code~\citep{kreber2021generating,cosler2023nl2spec,endres2024can,xie2025effective}.
While much of this work focuses on temporal properties such as linear temporal logic~(LTL),
recent studies demonstrate the promise of LLMs for generating method-level
postconditions.
However, evaluations are conducted on small, standalone programs,
limiting their ability to reflect real-world software complexity.

% \subsection{Specification Evaluation and Benchmarks}

Evaluating generated specifications remains difficult because there is no single
canonical ``correct and complete'' specification: multiple formulations may be
semantically valid yet differ substantially in behavioral coverage~\citep{endres2024can,xie2025effective}.
As a result, prior work frequently relies on manual evaluation or restricts
attention to tasks with well-defined ground truths~\citep{zhai2020c2s,kreber2021generating}.

Automatic evaluation of \emph{completeness} has received limited attention.
\citet{endres2024can} explore completeness checking for postconditions, but is limited to a small, Python-only dataset of standalone programs.
Existing benchmarks rarely capture repository-level code,
cross-language diversity, or the gap between test-based correctness and true
behavioral completeness.

In contrast, \projName~provides a multilingual, repository-level
benchmark with automatically evaluable ground-truth postconditions.
By combining test-based validation with mutant-based completeness checking,
it enables systematic measurement of both correctness and completeness in
realistic software settings.

\section{Methodology}
% This section outlines the benchmark design and curation pipeline, and defines how each instance is represented and evaluated. 
% We first describe the dataset structure and specification formats, and then detail the construction steps.

\subsection{Benchmark Overview}
\label{benchmark_overview}
\begin{figure*}[htbp] % h=here t=top b=bottom p=page of floats
  \centering
  \includegraphics[width=0.90\linewidth]{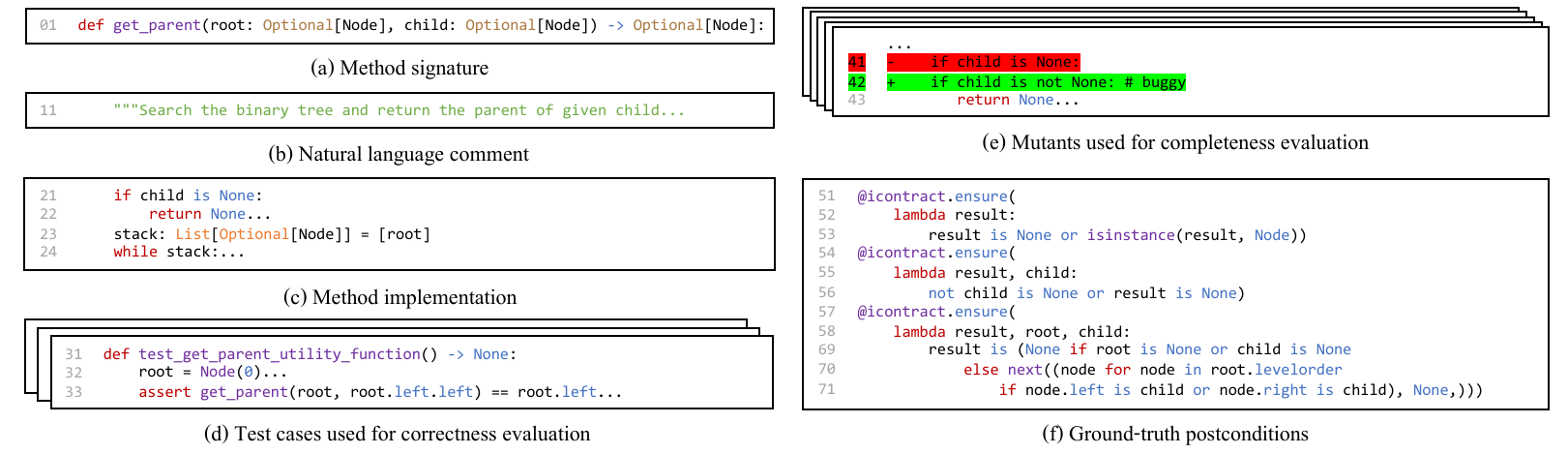}
  \caption{
  An example instance from \projName, illustrating the components of a benchmark instance. The instance corresponds to the target method \texttt{get\_parent}.
  }
  \label{fig:instance}
\end{figure*}

Our benchmark targets formal postcondition inference, evaluating whether approaches can generate precise and comprehensive postconditions from NL descriptions and/or code implementations.
We construct a multilingual, repository-level dataset with automated evaluation capabilities. 
Formally, each \projName~instance is a tuple that captures a method, its documentation, implementation, and evaluation artifacts:
\(x = \langle \text{sig}, \text{nl}, \text{impl}, \mathcal{T}, \mathcal{M}, \mathcal{P} \rangle\)
% \[
% x = \langle \text{sig}, \text{nl}, \text{impl}, \mathcal{T}, \mathcal{M}, \mathcal{P} \rangle .
% \] 
where \(\text{sig}\) is the method signature; \(\text{nl}\), the NL description; \(\text{impl}\), the code implementation; \(\mathcal{T}\), a set of test cases for correctness evaluation; \(\mathcal{M}\), a set of implementation mutants for completeness evaluation; and \(\mathcal{P}\), a set of ground-truth postconditions. 
\autoref{fig:instance} shows a representative instance example from \texttt{joowani/binarytree} \cite{binarytree-github}.%\juan{give citation, and give the consistent style when you mention a project name.}

Our postconditions are written in widely adopted formats, Java Modeling Language (JML~\cite{leavens2008jml}) for Java and icontract~\cite{icontract-github} for Python. 
Unlike \citet{endres2024can}, which uses native assertion statements as postconditions, and \citet{xie2025effective}, which uses a custom-defined format, the postcondition languages/frameworks used here are mature and more expressive. 

Our postconditions capture intra- and inter-class dependencies, and preserve pre-state information. 
For example, the postcondition \texttt{Arrays.deepEquals(\textbackslash old(this.items()), \textbackslash result.items())} uses \texttt{\textbackslash old(this.items())} to preserve the value of \texttt{this.items()} prior to method execution. 
Further, JML provides full support for logical connectives (implication, equivalence, and inequivalence; written as \(\Rightarrow\), \(\Leftrightarrow\), and \(\not\Leftrightarrow\))
While existing research highlights that these logical connectives improve writing and understanding \citep{Brown2024Implication, baron2024understanding}, \citet{endres2024can} and \citet{xie2025effective} do not provide sufficient support for them. 

This dataset is designed for flexibility and broad applicability, with a primary focus on specification generation. For generation from NL, we use \(x^{\text{genFromNL}} = (\text{sig}, \text{nl}, \mathcal{T}, \mathcal{M}, \mathcal{P})\); for inference from implementations, \(x^{\text{genFromCode}} = (\text{sig}, \text{impl}, \mathcal{T}, \mathcal{M}, \mathcal{P})\). Beyond specification tasks, the dataset also enables applications such as code generation using \(x^{\text{code}} = (\text{sig}, \text{nl}, \text{impl}, \mathcal{T}, \mathcal{P})\), which supports workflows that require grounding code synthesis in developer intent.

\subsection{Dataset Construction Pipeline}
\begin{figure*}[htbp] % h=here t=top b=bottom p=page of floats
  \centering
  \includegraphics[width=0.9\linewidth]{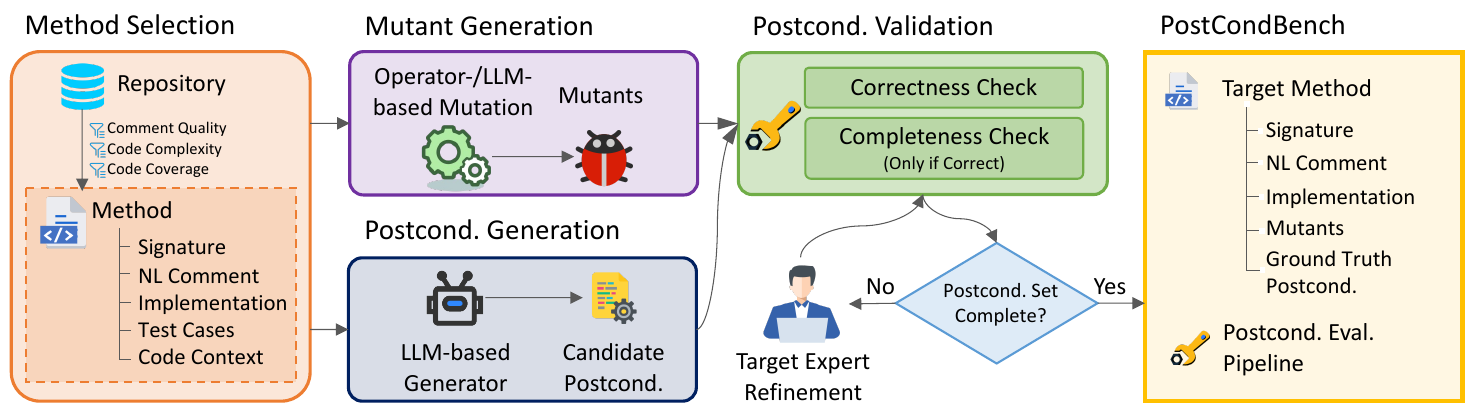}
  \caption{Overall workflow of \projName.}
  \label{fig:main_workflow}
\end{figure*}

We construct \projName~through a multi-step pipeline, illustrated in Figure~\ref{fig:main_workflow}.

\subsubsection{Target Method Selection}
We select candidate target methods from open-source repositories, requiring runnable environments and methods with adequate documentation, non-trivial complexity, and high test coverage.

\paragraph{Repository Selection.} We mine popular and diverse GitHub repositories: for Python/Java, we retrieve repos that (1) use a top-200 frequent tag, (2) are \(\leq 20\) MB, and (3) have a permissive license, then keep the top 5{,}000 most-starred per language. These sets cover all top-200 tags, ensuring broad topical diversity.

\paragraph{Automated Environment Setup.} To enable testing and coverage, each repository must run locally with tests passing and provide a coverage report~(e.g., JaCoCo~\cite{jacoco_github}, pytest-cov~\cite{pytestcov_github}). To scale the otherwise labor-intensive setup~(e.g., \citet{pan2025training}~reports over 200 hours spent on setup), we adopt an LLM-driven loop: starting from a clean machine with dependency managers (e.g., Poetry~\cite{poetry_github}, Maven~\cite{miller2010apache}), our module checks requirements and, upon failure, queries an LLM (\texttt{gpt-o4-mini}) to propose edits to core config files (e.g., \texttt{pyproject.toml}, \texttt{pom.xml}), applies them, and iterates for a bounded number of rounds. Out of 10{,}000 repositories, 941 are successfully set up (9.4\%), slightly better than SWE-agent~\cite{hu2025repo2run,yang2024swe}.

\paragraph{Method Parsing and Filtering.} We parse methods from the AST~\cite{latif2023comparison}, align coverage reports to obtain per-method line coverage, and retain methods that satisfy:
\ding{182} \textit{Comment Quality:} a comment with $>$15 English words (and we manually remove obvious mismatches);
\ding{183} \textit{Code Complexity:} at least 15 lines of code or cyclomatic complexity~\cite{mccabe1976complexity} \(\ge 3\);
\ding{184} \textit{Code Coverage:} \(\ge  90\%\) line coverage.
We exclude methods that throw exceptions under normal execution. This yields \textbf{1{,}030} Python and \textbf{1{,}476} Java methods.

\subsubsection{Mutant Generation}
\label{mutant_generation}
We generate mutants for each selected method following mutation-testing principles, which introduce small, controlled perturbations that approximate realistic bugs. This supports completeness evaluation: a complete postcondition set should be violated when the method runs on a defective implementation. We use operator-based mutation and LLM-based mutation, and keep only defective mutants that fail the original test suite.

\paragraph{Operator-based Mutation.} We apply a single, localized syntactic transformation to the original method body (e.g., replacing a return value with \texttt{None}, modifying conditional operators, or removing void method calls). For Python, we reuse mutation operators from Mutmut \cite{hovmoller2016mutmut}. For Java, we reproduce the default mutation operators from PIT \cite{coles2016pit}. 

\paragraph{LLM-based Mutation.} Operator-based mutation is efficient but limited in diversity and coupling strength~\cite{wang2024exploratory}. To complement it, we use an LLM-based mutation strategy. We parse the AST to locate lines that define conditions in \texttt{if}, \texttt{while}, and \texttt{do} statements, loop headers in \texttt{for} statements, and method calls. We replace each target line with a placeholder and prompt \texttt{gpt-4o-mini} to generate an alternative line that introduces a defect.

We discard mutants that still pass all tests and remove those that throw exceptions under normal execution, to ensure postconditions are exercised. We retain only methods with at least five valid mutants. This process yields 999 Python methods and 1{,}431 Java methods, with an average of 24.8 mutants per method.

\subsubsection{Postcondition Generation}
\label{postcond_generation}

We use an LLM to draft postcondition sets for each method, reducing manual effort in ground truth construction. 
The model is given (i) the target method, (ii) repository code, and (iii) a list of mutants, and is asked to propose postconditions that hold for the original code but fail for defective mutants.

We use \texttt{gpt-5-mini} with reasoning
%\footnote{https://platform.openai.com/docs/guides/reasoning} 
enabled. The prompt includes the method and, for each mutant, a line-by-line diff. We provide repository context by uploading a zip archive and using OpenAI's code interpreter tool \cite{openai-code-interpreter-docs}~to access relevant files.

% \subsubsection{Postcondition Generation}
% \label{postcond_generation}

% We use an LLM to draft postcondition sets for each method, providing strong candidates for the benchmark ground truths. To reduce manual refinement, the generator conditions on the target method, the full repository source code, and the mutant list. It is explicitly instructed to propose postconditions that hold for correct executions and are violated by defective mutants, aligning generation with our correctness and completeness criteria.

% In practice, we use \texttt{gpt-5-mini} with reasoning\footnote{https://platform.openai.com/docs/guides/reasoning} enabled. The prompt includes the method and the mutant list; for each mutant, we provide line-wise diffs against the original code. We supply the repository context by uploading a zip archive and invoking OpenAI's code interpreter tool\footnote{https://platform.openai.com/docs/guides/tools-code-interpreter}, which allows the model to parse the archive and retrieve any relevant code context during generation.

\subsubsection{Postcondition Validation}
\label{postcond_validation}
We validate each postcondition set along two axes: correctness on the original implementation and completeness against buggy mutants. A set is correct if the original method satisfies it across the provided tests; it is complete if it also distinguishes every generated mutant.

\paragraph{Correctness.} A postcondition set is correct if a correct implementation satisfies it for all (legal) inputs. We treat the developer test suite as the input distribution and require high test quality (line coverage \(\geq 90\%\)). For a target method \(m\) and a postcondition set \(p\), 
%where \(p\) contains at least one postcondition statement ~(e.g., \texttt{@icontract.ensure} or JML \texttt{ensures}), 
we instrument \(m\) with \(p\) and run the test suite, obtaining \(\eval(m,p)\in\{-1,0,1\}\): \(\eval(m,p)==1\) iff all tests pass and all postconditions hold; \(\eval(m,p)==0\) iff any postcondition is violated; and \(\eval(m,p)==-1\) iff the run fails due to other exceptions.
% let \(\eval(m,p) \in \{-1,0,1\}\) denote the test outcome when instrumenting \(m\) with \(p\):
% \begin{itemize}
%     \item \(\eval(m,p)=1\) if the test suite passes and all postconditions are satisfied.
%     \item \(\eval(m,p)=0\) if a postcondition is violated.
%     \item \(\eval(m,p)=-1\) if the run fails due to other exceptions. 
% \end{itemize}
We define correctness as
\[
\correct(m,p)==True \iff \eval(m,p)==1
\]
\begin{camchgblocktwo}
Our evaluation of correctness relies on test suites for the execution distribution.
To reduce noise from weak or sparse test suites, we filter tasks by test coverage and retain only methods with high line coverage.
Although high code coverage does not necessarily imply high test quality (e.g., assertions may be weak and corner cases may remain under-tested), which can lead to optimistic estimates in dynamic evaluation, we expect this criterion to preferentially select tasks with relatively stronger test suites in practice.
The final benchmark achieves an average line coverage of 99\%.
\end{camchgblocktwo}

\paragraph{Completeness.} Correctness alone can be vacuous; a postcondition set may pass tests yet fail to capture essential behavior. 
%For example, for a sorting routine, a postcondition that only checks output length can be correct but still misses the ordering property. 
We assess completeness via mutation testing. For each method \(m\), we generate a mutant set \(\{b_1,\dots,b_q\}\) (Section \ref{mutant_generation}).
For a correct postcondition set \(p\), we say \(p\) kills mutant \(b_h\) if \(\eval(b_h,p)==0\). A postcondition set is bug-complete if it is correct and kills all mutants:
\begin{empheq}{gather*}
\complete(m,p)==True 
\iff \\
\eval(m,p)==1 \land \left[\bigwedge_{h=1}^q \eval(b_h,p)==0 \right]
% l
\end{empheq}

\begin{camchgblocktwo}
This mutation-based completeness is an operational proxy for behavioral coverage rather than a definitive measure of semantic completeness.
Our mutation approach supports sufficient mutant quality and diversity, leading to a comprehensive and stable evaluation on completeness (see \autoref{sec:mutation_scheme_ablation}).
\end{camchgblocktwo}
%This validation yields an automated, test-based assessment of both correctness and completeness for each postcondition set.

\subsubsection{Expert Refinement and Finalization}
\label{expert_refinement}
If the LLM-generated postcondition set is not bug-complete, two domain experts refine it by inspecting the method, repository context, and mutants, and re-running validation until bug-completeness is reached; disagreements are resolved by consensus. Each expert has over five years of programming experience and more than one year of experience with formal specifications. If the generated set is already bug-complete, we adopt it as the ground truth. We exclude methods (or specific mutants) that require unsupported specification constructs or remain unkillable under the available tests.
All removals are reported in the \autoref{sec:expert_refinement}.
\camchg{Most of these exclusions arise from representational limits of the underlying specification frameworks.}

\paragraph{Benchmark Finalization.}
From 999 Python and 1{,}431 Java methods with sufficient mutants, we select a diverse subset via
farthest-first traversal~\cite{rosenkrantz1977analysis} within each language,
using cosine distance over CodeBERT~\cite{feng2020codebert} embeddings of method headers.
We then interleave automated generation with expert refinement and filtering,
yielding \projName~with 210 methods per language, each with a bug-complete postcondition set.

As shown in Table~\ref{tab:py_java_stats}, methods in \projName~exhibit 
non-trivial size and complexity, with rich test coverage and multiple ground-truth postconditions per method, reflecting realistic repository-level scenarios.

\begin{table}[htbp]
\centering
\caption{
Statistics of \projName~ for Python and Java.
All metrics are reported as method-level averages, except for the number of methods and repositories.
}
\label{tab:statistics}

\small
\setlength{\tabcolsep}{8pt}
\begin{tabular}{ccc}
\toprule
\textbf{Statistic} & \textbf{Python} & \textbf{Java} \\
\midrule
Methods & 210 & 210 \\
Repositories & 61  & 60  \\
Comment Word Length & 44.6 & 46.9 \\
Test Cases & 27.4 & 40.0 \\
Line Coverage & 98.8\% & 99.3\% \\
Lines of Code & 30.4 & 30.6 \\
Cyclomatic Complexity & 5.0 & 5.2 \\
\# Mutant & 26.6 & 21.9 \\
\# Postcondition & 3.6 & 5.5 \\
\bottomrule
\end{tabular}
\label{tab:py_java_stats}
\end{table}

\section{Experimental Setup 
% \juan{some details of the metrics etc. can be in the appendix.}
}
\label{sec:experiment-setup}

% Our experiments are conducted on \projName, a repository-level benchmark spanning Python and Java. 
% Each instance corresponds to a target method with accompanying code, NL documentation, tests, mutants, and ground-truth postconditions, as described in \cref{benchmark_overview}.

% The experimental setup is designed to answer the following research questions.

% \noindent
% \textbf{RQ1:} How well can LLMs generate correct and complete postconditions?

% \noindent \textbf{RQ2:} How large is the gap between test-based correctness and postcondition completeness? 
% \juan{I mean, these questions repeat each other. When you analyze their correctness and completeness, do not you just get the gaps? The same set of experiments? You should org them by experiments you did but the data you interpret.}

% \noindent \textbf{RQ3:} How do different input modalities (code, NL comments, and their combination) influence correctness and completeness?

% \noindent \textbf{RQ4:} What are the predominant failure modes of incorrect postconditions?

% \noindent \textbf{RQ5:} What behavioral constraints are most commonly missing from correct-but-incomplete postconditions?

% \noindent \textbf{RQ6:} How do method-level characteristics (e.g., method length and dependency complexity) relate to postcondition generation performance?

For each target method and each LLM, we evaluate postcondition generation under
three input settings. Under each setting, we independently sample 5 candidate postcondition sets. Each candidate is evaluated for test-based correctness and bug-completeness. Unless otherwise stated, we report single-sample~(\texttt{@1}) results in the main paper and defer multi-sample~ (\texttt{@k}) results to the appendix.

\paragraph{Input Settings.}

We evaluate postcondition generation under three input settings: 
\ding{182} Code-only~(C2P), which provides only the method implementation; \ding{183} NL-only~(N2P), which provides only NL comments; and \ding{184} Code+NL~(F2P), which provides both.
These settings allow us to study how different input modalities affect correctness, completeness, and their discrepancy.

\paragraph{Models.}

We evaluate 5 SOTA LLMs, covering proprietary and open-weight families: GPT-5~\cite{openai-gpt5}, Claude-4.5 (Sonnet)~\cite{anthropic-claude-4-5}, LLaMA-4 (Maverick)~\cite{llama-4}, Qwen3-32B~\cite{yang2025qwen3}, and Gemma-3-27B~\cite{mesnard2024gemma}. 
Details are in Appendix \ref{sec:models}.
%%
% All models are evaluated in a zero-shot setting with identical prompts, differing only in the provided input modality. For each model, method, and input setting, we independently generate 5 postcondition sets.
% %%
% For proprietary models, we enable their reasoning modes using the providers’ recommended configurations. For open-weight models, we deploy inference using \textsc{vLLM}~\cite{kwon2023efficient} with default settings.

\paragraph{Evaluation Metrics.}

We report Corr@k and Comp@k, which assess that when we generate $k$ postcondition sets for a method independently, the expectation that at least one generation result is correct/complete; full details are in Appendix~\ref{eval_metrics_app}.

\paragraph{Evaluation Protocol.}
For each model, input setting, and method, we evaluate generated postconditions using the metrics above.
We report model-level aggregates (e.g., Corr@1, Comp@1) in tables and method-level gap distributions in violin plots.

\section{Results}

We now present experimental results on \projName,
covering overall performance, the relationship between correctness and completeness,
the effect of different input modalities, and fine-grained analyses of model failures.

\subsection{How Large Is the Gap Between Correctness and Completeness?}

\begin{table*}[t]

\small

\setlength{\heavyrulewidth}{1.2pt}
\setlength{\lightrulewidth}{0.6pt}
\newcolumntype{Y}{>{\centering\arraybackslash}X}

\centering
\caption{
\camcaptiontwo{Postcondition generation performance of LLMs.}
\camcaptiontwo{We report Corr@k, Comp@k, $\Delta$@k, and Comp/Corr for $k \in \{1,3,5\}$, where Corr@k indicates the expectation that at least one of the $k$ generated postcondition sets is test-correct and Comp@k indicates for at least one is complete.}
\camcaptiontwo{Results are aggregated across input settings and programming languages.}
}
\label{tab:llm_results_new}

\centering
\begin{adjustbox}{max width=\textwidth}
\begin{tabular}{l|cccc|cccc|cccc}
\toprule
\multirow{3}{*}{\textbf{Model}} & \multicolumn{4}{c|}{\textbf{@1}} & \multicolumn{4}{c|}{\textbf{@3}} & \multicolumn{4}{c}{\textbf{@5}} \\
 & \multirow{2}{*}{\textbf{Corr}} & \multirow{2}{*}{\textbf{Comp}} & \multirow{2}{*}{\textbf{$\Delta$}} & \textbf{Comp/} & \multirow{2}{*}{\textbf{Corr}} & \multirow{2}{*}{\textbf{Comp}} & \multirow{2}{*}{\textbf{$\Delta$}} & \textbf{Comp/} & \multirow{2}{*}{\textbf{Corr}} & \multirow{2}{*}{\textbf{Comp}} & \multirow{2}{*}{\textbf{$\Delta$}} & \textbf{Comp/} \\
 & & & & \textbf{Corr} & & & & \textbf{Corr} & & & & \textbf{Corr} \\
\midrule
GPT-5 & 0.483 & \textbf{0.255} & 0.227 & \textbf{0.529} & 0.714 & \textbf{0.386} & 0.329 & \textbf{0.540} & 0.802 & \textbf{0.446} & 0.356 & \textbf{0.556} \\
Claude-4.5 & \textbf{0.629} & 0.207 & \textbf{0.423} & 0.329 & \textbf{0.777} & 0.268 & \textbf{0.509} & 0.345 & \textbf{0.822} & 0.292 & \textbf{0.530} & 0.355 \\
LLaMA-4 & 0.214 & 0.094 & 0.120 & 0.441 & 0.336 & 0.128 & 0.209 & 0.379 & 0.395 & 0.144 & 0.251 & 0.365 \\
Qwen3-32B & 0.118 & 0.055 & 0.063 & 0.466 & 0.189 & 0.080 & 0.109 & 0.423 & 0.229 & 0.093 & 0.136 & 0.406 \\
Gemma-3-27B & 0.110 & 0.044 & 0.066 & 0.401 & 0.171 & 0.058 & 0.113 & 0.339 & 0.204 & 0.064 & 0.140 & 0.315 \\
\bottomrule
\end{tabular}
\end{adjustbox}
\end{table*}

\begin{figure}[t] % h=here t=top b=bottom p=page of floats
  \centering
  \hspace*{-0.5cm}%
  \includegraphics[width=0.8\linewidth]{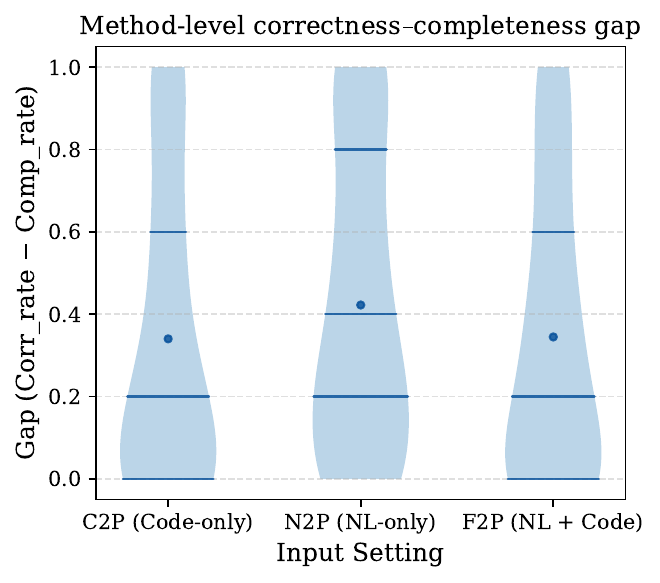}
  \caption{
  Distribution of method-level correctness–completeness gaps under different input settings.
  For each method $m$, the gap is computed as $\Delta(m) = \text{corr\_rate}(m) - \text{comp\_rate}(m)$, where corr\_rate and comp\_rate denote the fraction of generated postcondition sets that are test-correct and complete, respectively. 
  Each violin aggregates gap values across methods for which at least one test-correct postcondition is generated; methods with Corr@k = 0 are excluded.
  % Each violin aggregates gap values across all methods, pooling results from all models and both languages. 
  Horizontal lines indicate the 25th, 50th, and 75th percentiles, while dots denote the mean. 
  % The results show that the gap is widespread and varies systematically across input settings. 
  % The gap is most pronounced under the NL-only setting, indicating substantial overestimation by test-based correctness when relying solely on NL.
  }
  \label{fig:violin--main}
\end{figure}

\begin{camchgblock}
We evaluate LLM-generated postcondition sets under both single-sample and multi-sample settings.
\autoref{tab:llm_results_new} reports Corr@k and Comp@k, aggregated across languages and input settings.
Across all $k$, Claude-4.5 achieves the highest correctness, while GPT-5 achieves the strongest completeness.
% Completeness is consistently harder: even at $k=5$, the best model reaches Comp@5 of only 0.446, indicating that many test-correct outputs still fail to rule out buggy mutant behaviors.
\end{camchgblock}

To quantify this discrepancy, we measure the absolute gap $\Delta@1=\text{Corr@}1-\text{Comp@}1$ and the conditional completeness rate $\rho@1=\text{Comp@}1/\text{Corr@}1$.
Across models, $\Delta@1$ is large~(0.063–0.423) and $\rho@1$ is below 0.5 for most models, indicating fewer than half of test-correct outputs are complete.
Notably, higher Corr@1 can coincide with a larger gap~(e.g., Claude-4.5).

\begin{camchgblock}
\camcaptiontwo{Increasing the sample budget (i.e., increasing k in Corr/Comp@k metrics) improves both correctness and completeness, but does not remove the gap between them.
For Corr@k/Comp@k, GPT-5 improves from 0.483/0.255 at @1 to 0.802/0.446 at @5, while Claude-4.5 improves from 0.629/0.207 at @1 to 0.822/0.292 at @5.
}Additional sampling increases the chance of obtaining a usable specification. However, completeness still lags substantially behind correctness.
\end{camchgblock}

\autoref{fig:violin--main} analyzes method-level gaps, computed as
$\Delta(m)=\text{corr\_rate}(m)-\text{comp\_rate}(m)$ over multiple generations.
We focus on methods with $\text{Corr@}k>0$~(excluding $\text{Corr@}k=0$) to isolate incompleteness \emph{after}
correctness is achieved. The distribution is broad and right-skewed: the median gap is consistently positive~($\sim$0.2) and the upper quartile reaches roughly 0.6 to 0.8, showing the gap is widespread rather than driven by a
few outliers.

Input modality affects the \emph{severity} of the gap but not its existence.
All settings exhibit sizable gaps; even F2P has median $\sim$0.2 with a non-trivial tail above 0.5.
The gap is most pronounced under N2P, with the largest median and widest spread (many methods above 0.6,
extremes near 1.0). C2P and F2P are more concentrated; F2P slightly reduces the upper tail compared to C2P, but the
two remain close overall, suggesting code provides most of the constraining signal for completeness.

\begin{takeawaybox}
Correctness does not reliably imply completeness: large,
\camchg{systematic correctness--completeness gaps persist across methods, input settings, and sample budgets.}
NL-derived postconditions exhibit greater variance and heavier upper tails
\end{takeawaybox}

\subsection{What Are the Major Failure Modes of Incorrect Postconditions?}

\begin{figure*}[htbp] % h=here t=top b=bottom p=page of floats
  \centering
  \includegraphics[width=1\linewidth]{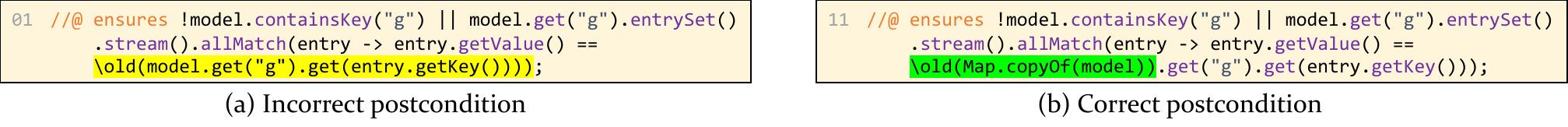}
  \caption{
  \camchgtwo{An example of incorrect postconditions caused by LLMs' misuse of specification languages or APIs, with corrected versions shown on the right. Highlighted regions indicate key corrections.
  % The inside-lambda variable \texttt{entry}~is created in the post-state, therefore cannot be involved in the snapshotted expression in \texttt{\textbackslash old}.
  }
  }
  \label{fig:limited_spec}
\end{figure*}

To understand why LLMs produce invalid specifications, we analyze failure modes of incorrect postconditions.
We sample 50 incorrect postcondition sets from GPT-5 and Claude-4.5; for each set, we select one postcondition that
triggers the failure and manually label its primary cause.

\paragraph{Specification Language / API Misuse (54\%).}
Most failures stem from incorrect contract syntax or framework APIs (54\%), leading to compilation/type errors or
runtime contract violations.

\camchgtwo{Note that many of these misuses are not superficial syntax mistakes. 
More than half of Java mistakes in this category are misusing post-state variables in \texttt{\textbackslash old($\cdot$)}.
In JML, the \texttt{\textbackslash old($\cdot$)} is used for snapshotting the value of the inputted expression at the pre-state. Expression that input to \texttt{\textbackslash old($\cdot$)} should only rely on variables that existed in the pre-state.
However, as an example in \cref{fig:limited_spec}, the generated postcondition uses \texttt{entry} in \texttt{\textbackslash old($\cdot$)}, while \texttt{entry} is a temporary variable that is created when calling the \texttt{allMatch($\cdot$)} API in the post-state.
Rather than merely a syntax error, this highlights an insufficient understanding of the boundary between pre-state and post-state.
% Another example in the upper part of the \cref{fig:limited_spec} also shows a mistake about the pre-state snapshotting: in \texttt{icontract}, pre-state snapshots require declaring an \texttt{OLD} parameter, but the model treats snapshot names as lambda parameters (``argument not set'').
}
% \cref{fig:limited_spec} shows two examples: in \texttt{icontract}, pre-state snapshots
% require declaring an \texttt{OLD} parameter, but the model treats snapshot names as lambda parameters (``argument not
% set''); in JML, \texttt{\textbackslash old($\cdot$)} is misused by referencing post-state variables inside it, although
% the expression must be fully pre-state.

\begin{camchgblock}
To test whether this failure mode is actionable, we add a brief grammar-guidance block to the F2P prompt, summarized from the official icontract and JML documentation, including an explanation of the specification language / API in use, with brief examples. We test GPT-5 and Claude-4.5.
\Cref{tab:grammar_guidance_f2p} shows that both correctness and completeness improve noticeably (e.g., 0.629 $\rightarrow$ 0.814 for Claude-4.5 Corr@1, and 0.207 $\rightarrow$ 0.283 for Comp@1).
However, the gap between correctness and completeness remains large.
Our claim still holds: even after these improvements, completeness remains challenging even for state-of-the-art models, and a large correctness--completeness gap is retained.
\end{camchgblock}

\paragraph{Semantic Overreach~(34\%).}
This error arises when LLMs enforce behaviors unsupported by the implementation or documentation.
For example, \cref{fig:hall_semantics} (Appendix) shows cases where LLMs formalize non-existent semantics.

\paragraph{Other Low-frequency Issues (12\%).}
The remaining cases include basic syntactic mistakes such as mismatched brackets (6\%)
and null dereferences (6\%), where postconditions access \texttt{None/null} values
without proper guards.

\begin{takeawaybox}
Incorrect postconditions are primarily caused by specification-language misuse
(54\%), indicating that improved model awareness of contract syntax and APIs
could eliminate a large fraction of failures.
% The second dominant source is semantic overreach (34\%),
% where models impose constraints not justified by the code or documentation.
\end{takeawaybox}

\subsection{Which Behaviors Are Missed by Correct but Incomplete Postconditions?}

We study postcondition sets that are test-correct on the reference implementation
but fail to kill at least one mutant, revealing missing behavioral constraints.
We focus on GPT-5 and Claude-4.5, and manually inspect 50 randomly sampled
(postcondition set, unkilled mutant) pairs.

\paragraph{Under-specified Return Value Behavior~(78\%).}
Most incompleteness arises from weak constraints on return values:
postconditions often validate only special cases, types, or coarse ranges,
leaving core computation insufficiently specified.
Consequently, output-perturbing or logic-bypassing mutants go undetected~(\cref{fig:corr_vs_comp} in Appendix).
Across the sampled cases, missed return-value behaviors span multiple return types:
scalar or primitive-like values (34\%), built-in collections of scalars (26\%),
repository-defined types (14\%), and third-party library types (4\%).
Notably, over one third of missed cases involve simple scalar returns,
indicating that incompleteness often stems from weak semantic constraints
rather than from output complexity.

\begin{figure}[htbp] % h=here t=top b=bottom p=page of floats
  \centering
  \includegraphics[width=1\linewidth]{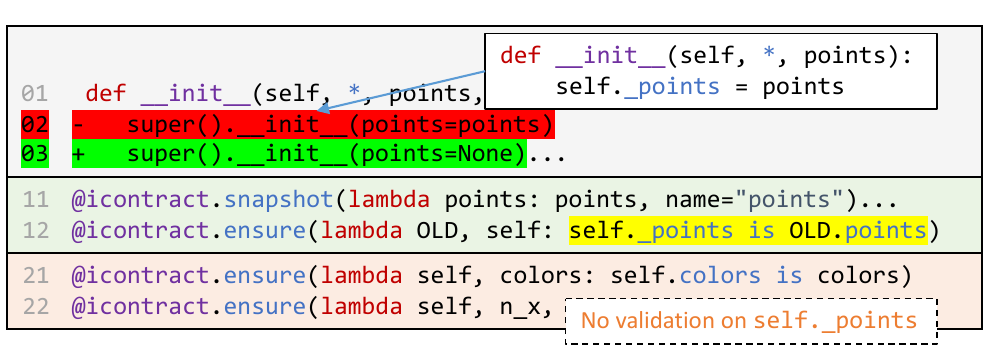}
\caption{Example of a correct but incomplete postcondition set.
Top: Target method with mutant changes~(``+''/``--'').
Middle: Complete postconditions.
Bottom: Test-correct but incomplete postconditions.}
  \label{fig:attribute}
\end{figure}

\paragraph{Unconstrained Object State~(16\%).}
Incompleteness also stems from missing constraints on object state.
As shown in \cref{fig:attribute}, while other fields are validated,
the initialization of \texttt{self.\_points} remains unconstrained,
allowing a state-altering mutant to pass undetected.

\paragraph{Lack of Defensive Guards~(6\%).}
Some postconditions lack basic robustness, raising runtime exceptions
(e.g., dereferencing \texttt{None}) under mutant-induced states rather than failing as specifications.
Adding simple defensive guards before attribute access would correctly reject these mutants.

\begin{takeawaybox}
Most incompleteness stems from weak constraints on return values,
often involving simple scalar outputs.
This suggests that systematic strengthening of return-value specifications,
rather than handling complex structures, could substantially improve completeness.
\end{takeawaybox}

\subsection{How Do Method Characteristics Affect Postcondition Quality?}
We analyze two method-level factors that affect postcondition generation:
dependency complexity (standalone vs.\ non-standalone) and method length.

\begin{table}[t]
\small
\setlength{\heavyrulewidth}{1.2pt}
\setlength{\lightrulewidth}{0.6pt}
\newcolumntype{Y}{>{\centering\arraybackslash}X}

\centering
\caption{
Postcondition generation performance on non-standalone (\textbf{Dep.}) and standalone (\textbf{Solo.}) methods.
\textbf{Dep.} requires repository-specific dependencies, while \textbf{Solo.} is runnable with only built-ins.
Bold indicates the higher score within each comparison.
}\label{tab:dependency_table--main}

\begin{tabularx}{0.4\textwidth}{cr|YY}
\toprule
{\textbf{Language}} &
{\textbf{Group}} &
{\textbf{Corr@1}} &
{\textbf{Comp@1}} \\

\midrule

\multirow{2}{*}{Python} & Dep.  & 0.181 & 0.035 \\
                        & Solo. & \textbf{0.239} & \textbf{0.090} \\

\midrule

\multirow{2}{*}{Java}   & Dep.  & 0.410 & 0.188 \\
                        & Solo. & \textbf{0.488} & \textbf{0.316} \\

\bottomrule
\end{tabularx}
\end{table}

\paragraph{Dependency Complexity.}
Following prior work~\citep{yu2024codereval}, we label a method as \emph{standalone}
if it runs with only built-in types and standard libraries; otherwise it is \emph{non-standalone}.
As shown in~\autoref{tab:dependency_table--main}, when averaging across models and input settings,
standalone methods consistently achieve higher scores.
The stratified results~(\autoref{tab:dependency_table_JZ} in Appendix) show this trend is robust:
across 60 paired comparisons~(2 metrics $\times$ 2 languages $\times$ 5 models $\times$ 3 settings),
standalone wins in 51 cases~(85\%).
The gap is larger for completeness: average Comp@1 rises from 0.035 to 0.090 in Python (2.6$\times$)
and from 0.188 to 0.316 in Java (1.7$\times$), suggesting that external dependencies
make it harder to produce sufficiently constraining postconditions.

\begin{figure}[htbp] % h=here t=top b=bottom p=page of floats
  \centering
  \includegraphics[width=1\linewidth]{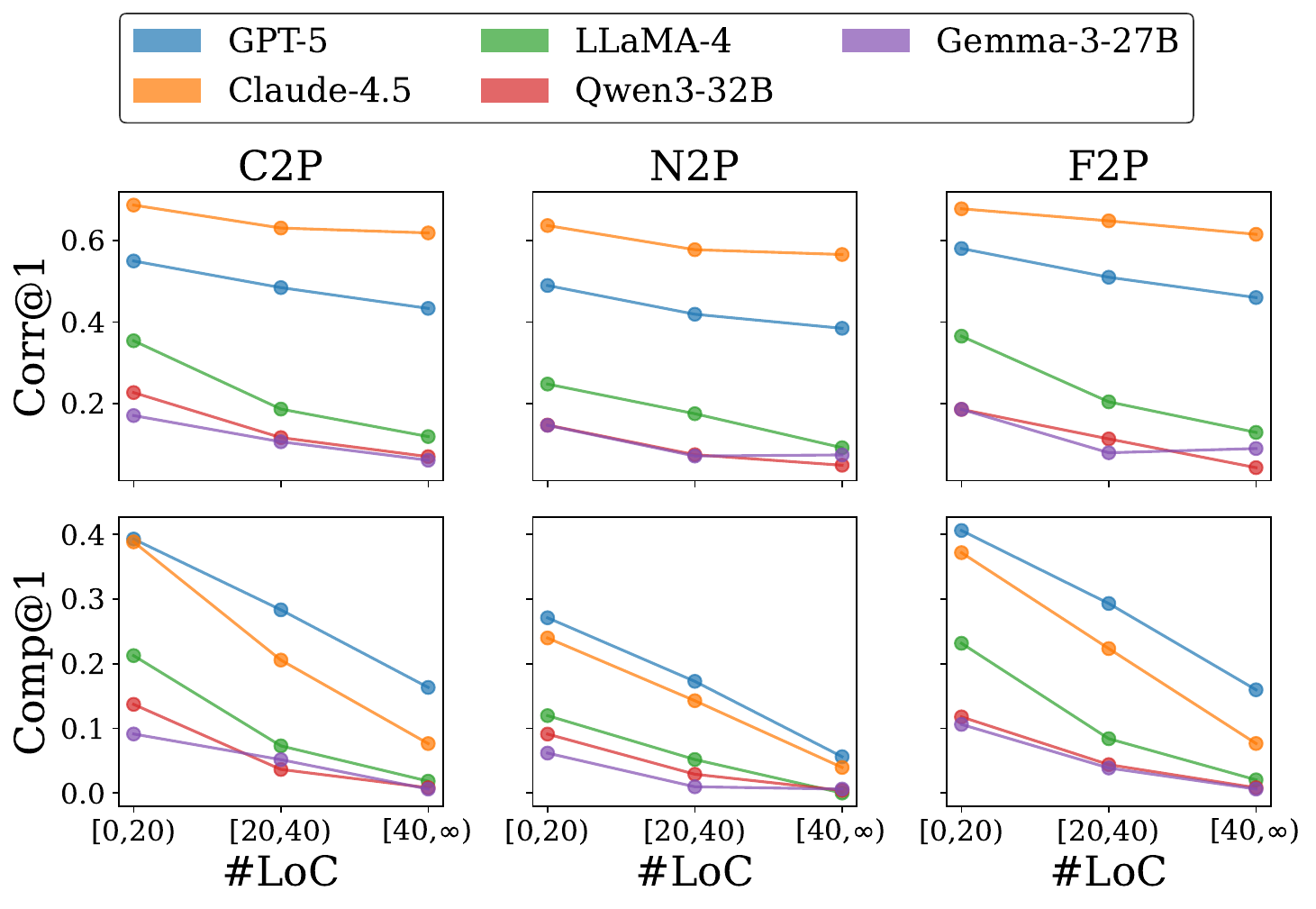}
  \caption{
Corr@1 (top) and Comp@1 (bottom) across target methods grouped by lines of code (LoC).
  }
  \label{fig:lines--main}
\end{figure}

\paragraph{Method Length.}
We bucket methods by lines of code (LoC): $[0,20)$, $[20,40)$, and $[40,\infty)$.
As shown in \autoref{fig:lines--main}, performance generally decreases with LoC,
and the drop is much steeper for completeness.
For example, under C2P with Claude-4.5, Comp@1 falls from 0.388 to 0.206 and 0.076
from short to medium and long methods.
Aggregated over models, languages, and input settings, the overall average score decreases
from 0.296 to 0.205 and then to 0.149 as LoC increases, confirming that longer methods
pose a substantially harder generation problem.

\begin{takeawaybox}
Method characteristics strongly affect postcondition generation:
standalone methods are easier than dependency-heavy ones, and longer methods degrade performance
monotonically, with completeness being especially sensitive.
\end{takeawaybox}

\subsection{\camchg{How Stable Is Completeness Under Mutant-Set Changes?}}

\begin{camchgblock}

We provide ablations to analyze how mutation design affects completeness (details in \autoref{sec:mutation_scheme_ablation}).

\paragraph{Operator-level ablation.}
Each language includes 11 operator-based mutation types. We remove each operator individually and compute Comp@1 (averaged across models). In Python, dropping any single operator changes Comp@1 by at most 0.004; in Java, most operators shift Comp@1 by $\leq$0.001 (see \autoref{tab:operator_ablation}).

\paragraph{Mutation-strength ablation.}
We further vary mutation ``strength'' by (i) randomly removing 5 of the 11 operators and (ii) randomly removing 50\% of LLM-based mutants (50 trials each). Across all models, the mean Comp@1 changes by less than 4\% relative to the full-mutant baseline, with standard deviation $\leq$0.004 in all settings (see \autoref{tab:mutation_stability}).

\paragraph{Mutant budget ablation.}
For each method, we randomly sample subsets containing 10\%, 20\%, $\dots$, 100\% of the full mutant set and compute Comp@1.
We observe a clear saturation pattern once $\geq$80\% of mutants are included. For GPT-5, mean Comp@1 decreases only slightly from 0.2616 (80\%) to 0.2585 (90\%) and 0.2551 (100\%), with absolute differences of $\approx$0.003. Claude-4.5 shows the same pattern (0.2120 $\rightarrow$ 0.2091 $\rightarrow$ 0.2068) (see \autoref{tab:mutation_stability}).

% \camchgtwo{In \projName, completeness is conditioned on the mutant set. To make this metric meaningful, the mutant pool must be both strong and diverse enough to cover a broad defect space. For this reason, we construct the set with complementary rule-based and LLM-based mutants rather than relying on a single mutation source.
% Our ablation experiments suggest that this design yields a reasonably stable metric in practice.
% We randomly remove 20\% of mutants and re-assess the completeness. A slight change in the completeness result is observed (e.g., 0.207 $\rightarrow$ 0.212 for Claude-4.5 Comp@1 and 0.094 $\rightarrow$ 0.097 for LLaMA-4 Comp@1).
% }Even under substantial perturbations, such as removing 5 of the 11 rule-based operators or removing 50\% of LLM-based mutants, the aggregate scores before and after the perturbations remain close to each other: for Claude-4.5, Comp@1 is 0.215 and 0.216 in the two settings, respectively, compared with the original Comp@1 0.207.
% These results suggest that our mutant set is sufficiently comprehensive to support a robust completeness evaluation. Detailed ablation results are provided in \autoref{sec:mutation_scheme_ablation}.
\end{camchgblock}

\begin{takeawaybox}
\camchg{
% Our completeness assessment depends on a strong and diverse mutant pool; it remains empirically stable in practice.
Comp@1 meaningfully reflects mutation design, remaining stable under substantial perturbations of operator sets and mutation budget. This indicates that the benchmark is neither dominated by a specific mutation family nor overly sensitive to particular operator choices.
}
\end{takeawaybox}

\section{Conclusion}

We presented \projName, a multilingual benchmark for evaluating method-level postconditions in real-world software. \projName~provides high-quality ground-truth postconditions and a runnable platform
that evaluates correctness with tests and completeness via defect discrimination.
Evaluating five SOTA LLMs across three generation tasks reveals a persistent gap between correctness and completeness.
% Evaluating five SOTA LLMs across three generation tasks reveals a persistent gap between correctness and completeness, which is further exacerbated by repository-level dependencies and method complexity.
By enabling reliable completeness evaluation, \projName~aims to support future work on generating more behaviorally complete postconditions.

\section*{Limitations}

\label{sec:limitations}

% \paragraph{Completeness relies on the mutant set.}
% Our completeness metric relies on the generated mutants.
% A postcondition set is considered more complete if it can distinguish the reference implementation from a broad range of mutated (buggy) variants.
% However, no practical mutation strategy can guarantee that the resulting mutants are exhaustive or fully representative of real bugs, so completeness measured this way may still miss behaviors outside the covered mutant space.
% To mitigate this limitation, we build on widely adopted mutation-testing toolchains to obtain strong, community-vetted rule-based mutants, and further augment them with a complementary LLM-based mutation scheme.
% By combining these two sources, we aim to cover a wider and more diverse defect space than either scheme alone, improving the robustness of completeness evaluation on top of well-established tooling.
% \camchg{Our analyses in \autoref{sec:mutation_scheme_ablation} further show that the metric is reasonably stable under substantial perturbations of the mutant set: operator- and LLM-based mutants are complementary, while completeness changes only modestly when we vary the mutant budget or randomly remove large subsets of mutation sources.}

\begin{camchgblocktwo}

% \paragraph{Correctness/completeness evaluation depends on tests.}
% Our evaluation of both correctness and completeness relies on test suites as the execution distribution.
% % \camchg{The benchmark should be interpreted as measuring correctness and completeness relative to the observed execution behavior rather than as a proof of full semantic correctness.}
% To reduce noise from weak or sparse test suites, we filter tasks by test coverage and retain only methods with high line coverage.
% Although high code coverage does not necessarily imply high test quality (e.g., assertions may be weak and corner cases may remain under-tested), which can lead to optimistic estimates in dynamic evaluation, we expect this criterion to preferentially select tasks with relatively stronger test suites in practice.
% The final benchmark achieves an average line coverage of 99\%.

We exclude methods (or specific mutants) that require unsupported specification constructs or remain unkillable under the available tests.
It is possible that this exclusion introduces bias in the \projName~evaluation of the affected behaviors, such as iterator or concurrency behaviors.
However, filtering affects a minority of the sampled methods. 
Most of these exclusions arise from representational limits of the underlying specification frameworks (\autoref{sec:expert_refinement}).
The exceptional behaviors are also not in \projName's scope since we focus on normal postconditions (in contrast to exceptional postconditions).
Besides, we restrict on methods with high code coverage due to the execution-based nature of our metrics.
\end{camchgblocktwo}

\section*{Ethical Considerations}

\paragraph{Data sources and licensing.}
Our benchmark is constructed from publicly available open-source repositories. We only include projects with permissive licenses and preserve attribution to the original authors and repositories. Any release of the benchmark and derived artifacts should comply with the licenses of the included projects.

\paragraph{Privacy and sensitive information.}
The benchmark is derived from public code and related artifacts (e.g., tests and comments). The dataset underwent end-to-end inspection during curation and validation. Across these inspections, we did not observe privacy-related or sensitive information in the data intended for release.

\paragraph{Human involvement.}
Some parts of dataset creation and validation involve domain-expert participation (e.g., refining or verifying postconditions and resolving ambiguities).
The experts are invited; no public recruitment. 
Domain experts provide assistance on a voluntary basis.
While administrative contact information may be required to coordinate participation, we did not collect such information as part of the dataset itself, did not use it beyond study administration, and it is not included in any released artifacts.

\section*{Acknowledgments}
This paper uses AI writing assistance purely with the language of the paper, which covers models used for paraphrasing or polishing the author's original content, rather than for suggesting new content.

\section*{Data Availability}\label{sec:data}
To follow the Open Science Policy and support reproducibility, we have released code about our implementations and evaluations.
\camchg{All source code and data used in our work can be found at~\url{https://github.com/zhang-ge-hao/postcond-bench}.}

\bibliography{bib/main, bib/rlwk, bib/rlwk2}

@misc{openai-code-interpreter-docs,
  author       = {OpenAI},
  title        = {Code Interpreter},
  howpublished = {\url{https://platform.openai.com/docs/guides/tools-code-interpreter}},
  note         = {OpenAI API Documentation. Accessed: 2026-01-06},
year={2023}
}

@misc{jsoup-jhy-github,
  author       = {jhy},
  title        = {jsoup: the Java HTML parser, built for HTML editing, cleaning, scraping, and XSS safety.},
  howpublished = {\url{https://github.com/jhy/jsoup/}},
  note         = {GitHub repository, accessed 2026-01-06},
   year={2025}
}

@misc{algorithms-keon-github,
  author       = {Keon},
  title        = {algorithms: Minimal examples of data structures and algorithms in Python and other languages},
  howpublished = {\url{https://github.com/keon/algorithms}},
  note         = {GitHub repository, accessed 2026-01-06},
  year={2025}
}

@misc{binarytree-github,
  author       = {Joowani},
  title        = {binarytree: A Python library for studying binary trees},
  howpublished = {\url{https://github.com/joowani/binarytree}},
  note         = {GitHub repository, accessed 2026-01-06},
  year={2022}
}

@misc{icontract-doc-async,
  author       = {{Parquery AG}},
  title        = {iContract Documentation: Async},
  howpublished = {\url{https://icontract.readthedocs.io/en/latest/async.html}},
  note         = {Accessed: 2026-01-05},
  year         = {2019}
}

@inproceedings{coles2016pit,
  title={Pit: a practical mutation testing tool for java},
  author={Coles, Henry and Laurent, Thomas and Henard, Christopher and Papadakis, Mike and Ventresque, Anthony},
  booktitle={Proceedings of the 25th international symposium on software testing and analysis},
  pages={449--452},
  year={2016}
}

@misc{hovmoller2016mutmut,
  author       = {Hovm{\"o}ller, Anders},
  title        = {Mutmut: a Python mutation testing system},
  howpublished = {\url{https://kodare.net/2016/12/01/mutmut-a-python-mutation-testing-system.html}},
  note         = {Blog post on En kodare},
  year         = {2016},
  urldate      = {2026-01-04}
}

@inproceedings{hu2025repo2run,
  title={Repo2run: Automated building executable environment for code repository at scale},
  author={Hu, Ruida and Peng, Chao and Xu, Junjielong and Gao, Cuiyun and others},
  booktitle={The Thirty-ninth Annual Conference on Neural Information Processing Systems},
  year={2025}
}

@book{miller2010apache,
  title={Apache Maven},
  author={Miller, Frederic P and Vandome, Agnes F and McBrewster, John},
  year={2010},
  publisher={Alpha Press}
}

@misc{poetry_github,
  title        = {{Poetry}: Python packaging and dependency management made easy},
  author       = {{python-poetry}},
  howpublished = {\url{https://github.com/python-poetry/poetry}},
  note         = {GitHub repository},
  year         = {2026},
  urldate      = {2026-01-04}
}

@misc{pytestcov_github,
  title        = {{pytest-cov}: Coverage Plugin for {pytest}},
  author       = {{pytest-dev}},
  howpublished = {\url{https://github.com/pytest-dev/pytest-cov}},
  note         = {GitHub repository},
  year         = {2025},
  urldate      = {2026-01-04}
}

@misc{jacoco_github,
  title        = {{JaCoCo}: Java Code Coverage Library},
  author       = {{jacoco}},
  howpublished = {\url{https://github.com/jacoco/jacoco}},
  note         = {GitHub repository},
  year         = {2025},
  urldate      = {2026-01-04}
}

@article{mccabe1976complexity,
  title={A complexity measure},
  author={McCabe, Thomas J},
  journal={IEEE Transactions on software Engineering},
  number={4},
  pages={308--320},
  year={1976},
  publisher={IEEE}
}

@inproceedings{latif2023comparison,
  title={Comparison of leading language parsers--antlr, javacc, sablecc, tree-sitter, yacc, bison},
  author={Latif, Afshan and Azam, Farooque and Anwar, Muhammad Waseem and Zafar, Amina},
  booktitle={2023 13th International Conference on Software Technology and Engineering (ICSTE)},
  pages={7--13},
  year={2023},
  organization={IEEE}
}

@misc{llama-4,
  author  = {Meta AI},
  title   = {GPT-4o mini: advancing cost-efficient intelligence},
  year    = {2025},
  month   = {4},
  day     = {5},
  url     = {https://ai.meta.com/blog/llama-4-multimodal-intelligence/},
  urldate = {2025-09-22},
  note    = {Meta}
}

@inproceedings{feng2020codebert,
  title={CodeBERT: A Pre-Trained Model for Programming and Natural Languages},
  author={Feng, Zhangyin and Guo, Daya and Tang, Duyu and Duan, Nan and Feng, Xiaocheng and Gong, Ming and Shou, Linjun and Qin, Bing and Liu, Ting and Jiang, Daxin and others},
  booktitle={Findings of the Association for Computational Linguistics: EMNLP 2020},
  pages={1536--1547},
  year={2020}
}

@article{rosenkrantz1977analysis,
  title={An analysis of several heuristics for the traveling salesman problem},
  author={Rosenkrantz, Daniel J and Stearns, Richard E and Lewis, II, Philip M},
  journal={SIAM journal on computing},
  volume={6},
  number={3},
  pages={563--581},
  year={1977},
  publisher={SIAM}
}

@article{wang2024exploratory,
  title={An Exploratory Study on Using Large Language Models for Mutation Testing},
  author={Wang, Bo and Chen, Mingda and Lin, Youfang and Papadakis, Mike and Zhang, Jie M},
  journal={CoRR},
  year={2024}
}

@inproceedings{jimenez2024swe,
  title={SWE-BENCH: CAN LANGUAGE MODELS RESOLVE REAL-WORLD GITHUB ISSUES?},
  author={Jimenez, Carlos E and Yang, John and Wettig, Alexander and Yao, Shunyu and Pei, Kexin and Press, Ofir and Narasimhan, Karthik},
  booktitle={12th International Conference on Learning Representations, ICLR 2024},
  year={2024}
}

@inproceedings{just2014mutants,
  title={Are mutants a valid substitute for real faults in software testing?},
  author={Just, Ren{\'e} and Jalali, Darioush and Inozemtseva, Laura and Ernst, Michael D and Holmes, Reid and Fraser, Gordon},
  booktitle={Proceedings of the 22nd ACM SIGSOFT international symposium on foundations of software engineering},
  pages={654--665},
  year={2014}
}

@article{jain2024livecodebench,
  title={Livecodebench: Holistic and contamination free evaluation of large language models for code},
  author={Jain, Naman and Han, King and Gu, Alex and Li, Wen-Ding and Yan, Fanjia and Zhang, Tianjun and Wang, Sida and Solar-Lezama, Armando and Sen, Koushik and Stoica, Ion},
  journal={arXiv preprint arXiv:2403.07974},
  year={2024}
}

@article{yang2024swe,
  title={Swe-agent: Agent-computer interfaces enable automated software engineering},
  author={Yang, John and Jimenez, Carlos E and Wettig, Alexander and Lieret, Kilian and Yao, Shunyu and Narasimhan, Karthik and Press, Ofir},
  journal={Advances in Neural Information Processing Systems},
  volume={37},
  pages={50528--50652},
  year={2024}
}

@misc{icontract-github,
  title   = {icontract},
  author  = {{Parquery}},
  url     = {https://github.com/Parquery/icontract},
  urldate = {2025-09-21},
  year    = {2025},
  note    = {GitHub repository}
}

@inproceedings{xie2025effective,
  title={How Effective are Large Language Models in Generating Software Specifications?},
  author={Xie, Danning and Yoo, Byoungwoo and Jiang, Nan and Kim, Mijung and Tan, Lin and Zhang, Xiangyu and Lee, Judy S},
  booktitle={2025 IEEE International Conference on Software Analysis, Evolution and Reengineering (SANER)},
  pages={1--12},
  year={2025},
  organization={IEEE}
}

@misc{leavens2008jml,
  title={JML reference manual},
  author={Leavens, Gary T and Poll, Erik and Clifton, Curtis and Cheon, Yoonsik and Ruby, Clyde and Cok, David and M{\"u}ller, Peter and Kiniry, Joseph and Chalin, Patrice and Zimmerman, Daniel M and others},
  year={2008}
}

@inproceedings{ma2025specgen,
  title={SpecGen: Automated Generation of Formal Program Specifications via Large Language Models},
  author={Ma, Lezhi and Liu, Shangqing and Li, Yi and Xie, Xiaofei and Bu, Lei},
  booktitle={2025 IEEE/ACM 47th International Conference on Software Engineering (ICSE)},
  pages={16--28},
  year={2025},
  organization={IEEE}
}

@inproceedings{baron2024understanding,
  title={Understanding Logical Expressions with Negations: Its Complicated},
  author={Baron, Aviad and Granot, Ilai and Yosef, Ron and Feitelson, Dror},
  booktitle={Proceedings of the 28th International Conference on Evaluation and Assessment in Software Engineering},
  pages={303--312},
  year={2024}
}

@techreport{Brown2024Implication,
  author      = {Walter E. Brown},
  title       = {Implication for C++},
  institution = {ISO/IEC JTC1/SC22/WG21},
  type        = {WG21 Paper P2971R2},
  number      = {P2971R2},
  date        = {2024-05-21},
  year        = {2024},
  url         = {https://www.open-std.org/jtc1/sc22/wg21/docs/papers/2024/p2971r2.pdf},
}

@inproceedings{
    pan2025training,
    title={Training Software Engineering Agents and Verifiers with {SWE}-Gym},
    author={Jiayi Pan and Xingyao Wang and Graham Neubig and Navdeep Jaitly and Heng Ji and Alane Suhr and Yizhe Zhang},
    booktitle={Forty-second International Conference on Machine Learning},
    year={2025},
    url={https://openreview.net/forum?id=Cq1BNvHx74}
}

@article{yang2025qwen3,
  title={Qwen3 technical report},
  author={Yang, An and Li, Anfeng and Yang, Baosong and Zhang, Beichen and Hui, Binyuan and Zheng, Bo and Yu, Bowen and Gao, Chang and Huang, Chengen and Lv, Chenxu and others},
  journal={arXiv preprint arXiv:2505.09388},
  year={2025}
}

@inproceedings{kwon2023efficient,
  title={Efficient memory management for large language model serving with pagedattention},
  author={Kwon, Woosuk and Li, Zhuohan and Zhuang, Siyuan and Sheng, Ying and Zheng, Lianmin and Yu, Cody Hao and Gonzalez, Joseph and Zhang, Hao and Stoica, Ion},
  booktitle={Proceedings of the 29th symposium on operating systems principles},
  pages={611--626},
  year={2023}
}

@misc{openai-gpt5,
  author  = {OpenAI},
  title   = {GPT-5 is here},
  year    = {2025},
  month   = {8},
  day     = {7},
  url     = {https://openai.com/gpt-5/},
  urldate = {2025-09-22},
  note    = {OpenAI}
}

@misc{anthropic-claude-4-5,
  author  = {Anthropic},
  title   = {Introducing Claude Sonnet 4.5},
  year    = {2025},
  month   = {9},
  day     = {29},
  url     = {https://www.anthropic.com/news/claude-sonnet-4-5},
  urldate = {2025-12-16},
  note    = {Anthropic}
}

@article{liu2023your,
  title={Is your code generated by chatgpt really correct? rigorous evaluation of large language models for code generation},
  author={Liu, Jiawei and Xia, Chunqiu Steven and Wang, Yuyao and Zhang, Lingming},
  journal={Advances in Neural Information Processing Systems},
  volume={36},
  pages={21558--21572},
  year={2023}
}

@inproceedings{islam2024mapcoder,
  title={MapCoder: Multi-Agent Code Generation for Competitive Problem Solving},
  author={Islam, Md Ashraful and Ali, Mohammed Eunus and Parvez, Md Rizwan},
  booktitle={Annual Meeting of the Association of Computational Linguistics 2024},
  pages={4912--4944},
  year={2024},
  organization={Association for Computational Linguistics (ACL)}
}

@inproceedings{xue2024llm4fin,
  title={Llm4fin: Fully automating llm-powered test case generation for fintech software acceptance testing},
  author={Xue, Zhiyi and Li, Liangguo and Tian, Senyue and Chen, Xiaohong and Li, Pingping and Chen, Liangyu and Jiang, Tingting and Zhang, Min},
  booktitle={Proceedings of the 33rd ACM SIGSOFT International Symposium on Software Testing and Analysis},
  pages={1643--1655},
  year={2024}
}

@inproceedings{lamsweerde2000formal,
  title={Formal specification: a roadmap},
  author={Lamsweerde, Axel van},
  booktitle={Proceedings of the Conference on the Future of Software Engineering},
  pages={147--159},
  year={2000}
}

@article{mesnard2024gemma,
  title        = {Gemma: Open Models Based on Gemini Research and Technology},
  author       = {Gemma Team Thomas Mesnard and Cassidy Hardin and Robert Dadashi and Surya Bhupatiraju and Shreya Pathak and L. Sifre and Morgane Riviere and Mihir Kale and J Christopher Love and Pouya Dehghani Tafti and L'eonard Hussenot and Aakanksha Chowdhery and Adam Roberts and Aditya Barua and Alex Botev and Alex Castro-Ros and Ambrose Slone and Am'elie H'eliou and Andrea Tacchetti and Anna Bulanova and Antonia Paterson and Beth Tsai and Bobak Shahriari and Charline Le Lan and Christopher A. Choquette-Choo and Cl'ement Crepy and Daniel Cer and Daphne Ippolito and David Reid and Elena Buchatskaya and Eric Ni and Eric Noland and Geng Yan and George Tucker and George-Christian Muraru and Grigory Rozhdestvenskiy and Henryk Michalewski and Ian Tenney and Ivan Grishchenko and Jacob Austin and James Keeling and Jane Labanowski and Jean-Baptiste Lespiau and Jeff Stanway and Jenny Brennan and Jeremy Chen and Johan Ferret and Justin Chiu and Justin Mao-Jones and Katherine Lee and Kathy Yu and Katie Millican and Lars Lowe Sjoesund and Lisa Lee and Lucas Dixon and Machel Reid and Maciej Mikula and Mateo Wirth and Michael Sharman and Nikolai Chinaev and Nithum Thain and Olivier Bachem and Oscar Chang and Oscar Wahltinez and Paige Bailey and Paul Michel and Petko Yotov and Pier Giuseppe Sessa and Rahma Chaabouni and Ramona Comanescu and Reena Jana and Rohan Anil and Ross McIlroy and Ruibo Liu and Ryan Mullins and Samuel L Smith and Sebastian Borgeaud and Sertan Girgin and Sholto Douglas and Shree Pandya and Siamak Shakeri and Soham De and Ted Klimenko and Tom Hennigan and Vladimir Feinberg and Wojciech Stokowiec and Yu-hui Chen and Zafarali Ahmed and Zhitao Gong and Tris Brian Warkentin and Ludovic Peran and Minh Giang and Cl'ement Farabet and Oriol Vinyals and Jeffrey Dean and Koray Kavukcuoglu and Demis Hassabis and Zoubin Ghahramani and Douglas Eck and Joelle Barral and Fernando Pereira and Eli Collins and Armand Joulin and Noah Fiedel and Evan Senter and Alek Andreev and Kathleen Kenealy},
  year         = 2024,
  journal      = {ArXiv},
  volume       = {abs/2403.08295},
  url          = {https://api.semanticscholar.org/CorpusID:268379206}
}

@inproceedings{cosler2023nl2spec,
  title        = {nl2spec: Interactively translating unstructured natural language to temporal logics with large language models},
  author       = {Cosler, Matthias and Hahn, Christopher and Mendoza, Daniel and Schmitt, Frederik and Trippel, Caroline},
  year         = 2023,
  booktitle    = {International Conference on Computer Aided Verification},
  pages        = {383--396},
  organization = {Springer}
}

@article{kreber2021generating,
  title={Generating symbolic reasoning problems with transformer gans},
  author={Kreber, Jens U and Hahn, Christopher},
  journal={arXiv preprint arXiv:2110.10054},
  year={2021}
}

@inproceedings{xie2022docter,
  title        = {DocTer: documentation-guided fuzzing for testing deep learning API functions},
  author       = {Xie, Danning and Li, Yitong and Kim, Mijung and Pham, Hung Viet and Tan, Lin and Zhang, Xiangyu and Godfrey, Michael W},
  year         = 2022,
  booktitle    = {Proceedings of the 31st ACM SIGSOFT International Symposium on Software Testing and Analysis},
  pages        = {176--188}
}

@article{chen2021evaluating,
  title        = {Evaluating Large Language Models Trained on Code},
  author       = {Mark Chen and Jerry Tworek and Heewoo Jun and Qiming Yuan and Henrique Ponde and Jared Kaplan and Harrison Edwards and Yura Burda and Nicholas Joseph and Greg Brockman and Alex Ray and Raul Puri and Gretchen Krueger and Michael Petrov and Heidy Khlaaf and Girish Sastry and Pamela Mishkin and Brooke Chan and Scott Gray and Nick Ryder and Mikhail Pavlov and Alethea Power and Lukasz Kaiser and Mohammad Bavarian and Clemens Winter and Philippe Tillet and Felipe Petroski Such and David W. Cummings and Matthias Plappert and Fotios Chantzis and Elizabeth Barnes and Ariel Herbert-Voss and William H. Guss and Alex Nichol and Igor Babuschkin and Suchir Balaji and Shantanu Jain and Andrew Carr and Jan Leike and Joshua Achiam and Vedant Misra and Evan Morikawa and Alec Radford and Matthew M. Knight and Miles Brundage and Mira Murati and Katie Mayer and Peter Welinder and Bob McGrew and Dario Amodei and Sam McCandlish and Ilya Sutskever and Wojciech Zaremba},
  year         = 2021,
  journal      = {ArXiv},
  volume       = {abs/2107.03374},
  url          = {https://api.semanticscholar.org/CorpusID:235755472}
}

@inproceedings{yu2024codereval,
  title={Codereval: A benchmark of pragmatic code generation with generative pre-trained models},
  author={Yu, Hao and Shen, Bo and Ran, Dezhi and Zhang, Jiaxin and Zhang, Qi and Ma, Yuchi and Liang, Guangtai and Li, Ying and Wang, Qianxiang and Xie, Tao},
  booktitle={Proceedings of the 46th IEEE/ACM International Conference on Software Engineering},
  pages={1--12},
  year={2024}
}

@inproceedings{zhai2020c2s,
  title        = {C2S: translating natural language comments to formal program specifications},
  author       = {Zhai, Juan and Shi, Yu and Pan, Minxue and Zhou, Guian and Liu, Yongxiang and Fang, Chunrong and Ma, Shiqing and Tan, Lin and Zhang, Xiangyu},
  year         = 2020,
  booktitle    = {Proceedings of the 28th ACM Joint Meeting on European Software Engineering Conference and Symposium on the Foundations of Software Engineering},
  pages        = {25--37}
}

@inproceedings{zhang2020automated,
  title        = {Automated geration of LTL Specifications For Smart Home IoT Using Natural Language},
  author       = {Zhang, Shiyu and Zhai, Juan and Lei, Bu and Linzhang, Wang and Xuandong Li},
  year         = 2020,
  booktitle    = {2020 Design, Automation \& Test in Europe Conference \& Exhibition (DATE)},
  organization = {IEEE}
}

@inproceedings{blasi2018translating,
  title        = {Translating code comments to procedure specifications},
  author       = {Blasi, Arianna and Goffi, Alberto and Kuznetsov, Konstantin and Gorla, Alessandra and Ernst, Michael D and Pezz{\`e}, Mauro and Castellanos, Sergio Delgado},
  year         = 2018,
  month        = jul,
  journal      = {Proceedings of the 27th ACM SIGSOFT International Symposium on Software Testing and Analysis},
  booktitle    = {Proceedings of the 27th ACM SIGSOFT International Symposium on Software Testing and Analysis},
  address      = {Amsterdam, Netherlands},
  pages        = {242--253},
  url          = {https://api.semanticscholar.org/CorpusID:49863340},
  organization = {ACM}
}

@inproceedings{goffi2016automatic,
  title        = {Automatic generation of oracles for exceptional behaviors},
  author       = {Goffi, Alberto and Gorla, Alessandra and Ernst, Michael D and Pezz{\`e}, Mauro},
  year         = 2016,
  booktitle    = {Proceedings of the 25th International Symposium on Software Testing and Analysis},
  pages        = {213--224},
  organization = {ACM}
}

@inproceedings{nguyen2014mining,
  title        = {Mining preconditions of APIs in large-scale code corpus},
  author       = {Nguyen, Hoan Anh and Dyer, Robert and Nguyen, Tien N and Rajan, Hridesh},
  year         = 2014,
  booktitle    = {Proceedings of the 22nd ACM SIGSOFT International Symposium on Foundations of Software Engineering},
  pages        = {166--177},
  organization = {ACM}
}

@inproceedings{pandita2012inferring,
  title        = {Inferring method specifications from natural language {API} descriptions},
  author       = {Pandita, Rahul and Xiao, Xusheng and Zhong, Hao and Xie, Tao and Oney, Stephen and Paradkar, Amit},
  year         = 2012,
  journal      = {2012 34th International Conference on Software Engineering (ICSE)},
  booktitle    = {Proceedings of the 34th International Conference on Software Engineering},
  pages        = {815--825},
  url          = {https://api.semanticscholar.org/CorpusID:7449460},
  organization = {IEEE Press}
}

@article{henkel2008developing,
  title        = {Developing and debugging algebraic specifications for Java classes},
  author       = {Henkel, Johannes and Reichenbach, Christoph and Diwan, Amer},
  year         = 2008,
  journal      = {ACM Transactions on Software Engineering and Methodology (TOSEM)},
  publisher    = {ACM New York, NY, USA},
  volume       = 17,
  number       = 3,
  pages        = {1--37}
}

@inproceedings{ramanathan2007static,
  title        = {Static specification inference using predicate mining},
  author       = {Ramanathan, Murali Krishna and Grama, Ananth and Jagannathan, Suresh},
  year         = 2007,
  booktitle    = {ACM SIGPLAN Notices},
  pages        = {123--134},
  organization = {ACM}
}

@article{nimmer2002automatic,
  title        = {Automatic generation of program specifications},
  author       = {Nimmer, Jeremy W and Ernst, Michael D},
  year         = 2002,
  journal      = {ACM SIGSOFT Software Engineering Notes},
  publisher    = {ACM New York, NY, USA},
  volume       = 27,
  number       = 4,
  pages        = {229--239}
}

@article{ernst2001dynamically,
  title        = {Dynamically discovering likely program invariants to support program evolution},
  author       = {Ernst, Michael D and Cockrell, Jake and Griswold, William G and Notkin, David},
  year         = 2001,
  journal      = {IEEE Transactions on Software Engineering},
  publisher    = {IEEE},
  volume       = 27,
  number       = 2,
  pages        = {99--123}
}

@phdthesis{snook2001exploring,
  title        = {Exploring the barriers to formal specification},
  author       = {Snook, Colin Frank},
  year         = 2001,
  school       = {University of Southampton}
}

@article{Chen2016SupportingOC,
  title        = {Supporting oracle construction via static analysis},
  author       = {Junjie Chen and Yanwei Bai and Dan Hao and Lingming Zhang and Lu Zhang and Bing Xie and Hong Mei},
  year         = 2016,
  journal      = {2016 31st IEEE/ACM International Conference on Automated Software Engineering (ASE)},
  pages        = {178--189},
  url          = {https://api.semanticscholar.org/CorpusID:472942}
}

@inproceedings{Flanagan2001HoudiniAA,
  title        = {Houdini, an Annotation Assistant for ESC/Java},
  author       = {Cormac Flanagan and K. Rustan M. Leino},
  year         = 2001,
  booktitle    = {FME},
  url          = {https://api.semanticscholar.org/CorpusID:1534849}
}

@article{Shoham2007StaticSM,
  title        = {Static Specification Mining Using Automata-Based Abstractions},
  author       = {Sharon Shoham and Eran Yahav and Stephen J. Fink and Marco Pistoia},
  year         = 2007,
  journal      = {IEEE Transactions on Software Engineering},
  volume       = 34,
  pages        = {651--666},
  url          = {https://api.semanticscholar.org/CorpusID:2483401}
}

@article{endres2024can,
  author    = {Endres, Madeline and Fakhoury, Sarah and Chakraborty, Saikat and Lahiri, Shuvendu K},
  journal   = {Proceedings of the ACM on Software Engineering},
  number    = {FSE},
  pages     = {1889-1912},
  publisher = {ACM New York, NY, USA},
  title     = {Can large language models transform natural language intent into formal method postconditions?},
  volume    = {1},
  year      = {2024}
}

@inproceedings{just2014defects4j,
  title={Defects4J: A database of existing faults to enable controlled testing studies for Java programs},
  author={Just, Ren{\'e} and Jalali, Darioush and Ernst, Michael D},
  booktitle={Proceedings of the 2014 international symposium on software testing and analysis},
  pages={437--440},
  year={2014}
}

\clearpage

\appendix

\newpage

\begin{camchgblocktwo}

\section{Discussion}

\paragraph{Insufficiency for Real Bugs.}
We use mutants rather than real bugs made in development to assess the completeness. The reason is that mutation lets us generate many defective variants for the same method.
For real bugs \cite{just2014defects4j, jimenez2024swe}, each fixed version is typically paired with a very limited number (usually only one) of corresponding buggy versions, which makes it insufficient to assess how completely a specification constrains the method's behavior.
% Second, prior work shows that mutants can effectively couple with real faults~\cite{just2014mutants}.
% However, we also admit that real bugs and mutants are complementary rather than interchangeable. Extending the benchmark with real bugs will be meaningful future work.

Moreover, prior work shows that mutants can effectively couple with real faults~\cite{just2014mutants}.
Therefore, although \projName~is designed to evaluate the quality of formal postconditions, the reported mutant discrimination can be a prerequisite capability for downstream uses such as verification and debugging.

\section{Future Work}

\paragraph{Live Benchmark against Data Leakage.}
``Live'' benchmarks are continuously updated with newer data, reducing the risk of data leakage \cite{jain2024livecodebench}. They are often designed to support efficient updates.
For \projName, most stages of the construction pipeline are automated, and the benchmark can be re-run on newly released repositories with limited manual effort; the expert effort for maintaining a refreshed version is expected to be less than 50 hours per year. It is practical to maintain a periodically refreshed benchmark.
% The same property also makes the benchmark easier to expand than prior datasets that rely heavily on manual curation and evaluation.

% \paragraph{Beyond Normal Postconditions.}
% The current benchmark focuses on normal postconditions; in the future, incorporating additional evaluation mechanisms to assess exceptional postconditions, preconditions, and invariants will be valuable work.

% \paragraph{Include More Behaviors.}
% We also apply moderate data filtering during benchmark construction, which would cause underrepresented behaviors such as iterator-sensitive logic and language-entity manipulation. More details in \autoref{sec:expert_refinement}. 
% We view richer support for such cases as an important direction for future benchmark expansion.

\end{camchgblocktwo}

\newcolumntype{L}{>{\raggedright\arraybackslash}p{0.22\textwidth}}
\newcolumntype{C}{>{\centering\arraybackslash}p{0.12\textwidth}}
\newcolumntype{O}{>{\centering\arraybackslash}p{0.05\textwidth}}

\begin{table*}[t]
\centering
\caption{Comparison of evaluation features in Python/Java postcondition generation.
% \juan{check whether the content in my table is correct or not, i roughly checked and it seems to be correct}
} 
\label{tab:work_compare_JZ}
\small
\setlength{\tabcolsep}{3pt}
\renewcommand{\arraystretch}{1.12}

\begin{threeparttable}
\begin{tabularx}{\textwidth}{@{}L|CCCCCO@{}}
\toprule
\textbf{Work} &
\textbf{C2S \citep{zhai2020c2s}} &
\textbf{\citet{endres2024can} (HumanEval+)} &
\textbf{\citet{endres2024can} (Defects4J)} &
\textbf{SpecGen \citep{ma2025specgen}} &
\textbf{\citet{xie2025effective}} &
\textbf{Ours} \\
\midrule

Multi-Language            & \xmark & \xmark & \xmark & \xmark & \xmark & \cmark \\
Repository-level          & \cmark & \xmark & \cmark & \xmark & \cmark & \cmark \\
Ground Truths Included    & \xmark & \xmark & \xmark & \cmark & \cmark & \cmark \\
Advanced Spec. Syntax     & \cmark & \xmark & \xmark & \cmark & \xmark & \cmark \\
Auto.\ Completeness Eval. & \xmark & \cmark & \xmark & \xmark & \xmark & \cmark \\

\bottomrule
\end{tabularx}

\begin{tablenotes}[flushleft]
\footnotesize
\item \citet{endres2024can} report postcondition generation results under two benchmarks:
\texttt{HumanEval+}~\citep{liu2023your} and \texttt{Defects4J}~\citep{just2014defects4j}.
\end{tablenotes}
\end{threeparttable}
\end{table*}

% \juan{I don't know why i have two identical tables. here we have table 3, should be table 1. fix it.}
\section{Comparison with Existing Work}
\cref{tab:work_compare_JZ} positions \projName~relative to prior work on postcondition generation.
While existing studies primarily propose generation approaches and evaluate them under task-specific datasets or execution settings, they do not release a reusable benchmark with curated ground truths or an explicit evaluation methodology for completeness.
\projName~fills this gap by providing a benchmark that jointly supports multilingual, repository-level programs, advanced postcondition syntax, and automatic evaluation of both correctness and completeness.

\section{Experimental Setup}

\label{sec:experiment-setup-detail}

\subsection{Models}
\label{sec:models}

We chose five state-of-the-art large language models (LLMs) for experiments: GPT-5 \cite{openai-gpt5}, Claude-4.5 (Sonnet) \cite{anthropic-claude-4-5}, LLaMA-4 (Maverick) \cite{llama-4}, Qwen3-32B \cite{yang2025qwen3}, Gemma-3-27B \cite{mesnard2024gemma}.
\camchg{For GPT-5, we use the OpenAI batch API.
For Claude-4.5 and LLaMA-4, we use the batch inference pipeline in Amazon Bedrock.}
For Qwen and Gemma models, we locally deploy Vllm \cite{kwon2023efficient}~services for experiments. 
We apply the reasoning mode for GPT-5 and Claude 4.5, setting \texttt{reasoning.effort = "medium"}~for GPT-5 and \texttt{thinking.budget\_tokens = 2048}~for Claude 4.5.
\camchg{Besides the reasoning effort configurations, we follow each provider's default configurations and use consistent prompts across models for fair comparison.}

\begin{figure}[htbp] % h=here t=top b=bottom p=page of floats
  \centering
  \includegraphics[width=1\linewidth]{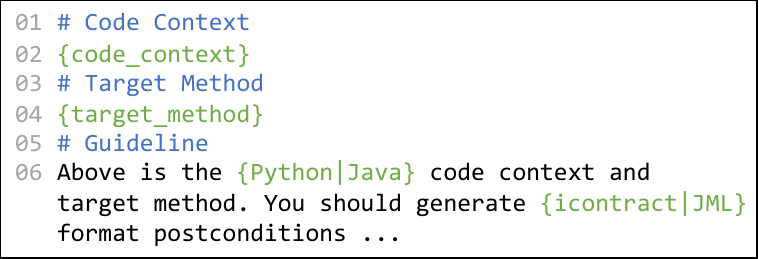}
  \caption{Prompt template. The placeholders are shown in green.}
  \label{fig:prompt}
\end{figure}

\subsection{Prompt Construction}

\projName~supports runnable repositories and automatic evaluation; we therefore formulate postcondition generation as prompting an LLM with a target method plus its surrounding context.
All settings share the prompt template in \cref{fig:prompt}, which contains two main fields: \texttt{\{code\_context\}} (in-file context) and \texttt{\{target\_method\}} (the method to validate).

We instantiate three input settings:
\begin{itemize}
    \item \textbf{F2P.}
    We provide the method signature, the NL method-level comment, and the full implementation in \texttt{\{target\_method\}} (cf.\ \cref{fig:instance}).
    We fill \texttt{\{code\_context\}} with the method-located file content to provide in-file context.
    \item \textbf{C2P.}
    We provide the method signature and implementation in \texttt{\{target\_method\}}, removing any code comments inside the implementation.
    We also remove all code comments from \texttt{\{code\_context\}}.
    \item \textbf{N2P.}
    We provide the method signature and its NL method-level comment in \texttt{\{target\_method\}}.
    In \texttt{\{code\_context\}}, we keep the method-located file but remove method bodies, so the model can leverage declarations and surrounding structure without seeing implementations.
\end{itemize}
These settings isolate the effects of code, comments, and their combination on correctness and bug-completeness.

% Various tasks can be formulated from \projName~since our benchmark introduces runnable full repositories and an automatic evaluation pipeline.
% We formulate three generation tasks. 
% They share the prompt format shown in \cref{fig:prompt}.
% \begin{itemize}
%     \item Full Original Context Inputted (F2P): 
%     We concatenate the method signature, natural language (NL) method-level comment, and implementation (sub-figures (a), (b), and (c) in \cref{fig:instance}) to fulfill the \texttt{\{method\}}~placeholder. 
%     For \texttt{\{code\_context\}}~placeholder, we use the method-located Python/Java file content to provide in-file context.
%     %
%     \item Pure Code Inputted (C2P): 
%     The method signature and implementation fulfill the \texttt{\{method\}}~placeholder, with the code comments inside the implementation also removed.
%     For \texttt{\{code\_context\}}~placeholder, the method-located file content will be added without any code comments.
%     %
%     \item NL Comment Inputted (N2P):
%     The method signature and NL method-level comment fulfill the \texttt{\{method\}}~placeholder.
%     For the \texttt{\{code\_context\}}~placeholder, the method-located file content will be added, but all method implementations are removed.
% \end{itemize}
% These three settings can help us explore how code, comments, and their combination affect correctness and completeness.

% \input{src/exp_setting/metrics}

\subsection{Evaluation Metrics}
\label{eval_metrics_app}

We evaluate generated postconditions along two dimensions: correctness and completeness.
Completeness is evaluated only for test-correct postcondition sets; candidates that fail correctness are treated as incomplete.

\paragraph{Correctness.}
A generated \emph{postcondition set} is considered \emph{correct} if, when instrumented into the original method, all postconditions in the set pass all provided test cases. 
Test-based correctness reflects whether the generated specification is consistent with observed program behaviors.

\paragraph{Completeness.}
A generated \emph{postcondition set} is considered \emph{complete} if it not only passes all test cases, but also rejects all automatically generated mutants of the target method. Since mutants introduce buggy behaviors that should be ruled out by a correct specification, completeness measures whether a postcondition set sufficiently constrains the method’s behavior. Completeness is evaluated only for test-correct postcondition sets; candidates that fail correctness are treated as incomplete.

\noindent
\(\bullet\) {\texttt{@k} aggregation.}
Let $k$ denote the number of independently generated postcondition sets for a given method.
\textbf{Corr@k} indicates whether at least one of the $k$ generated postcondition sets is test-correct.
\textbf{Comp@k} indicates whether at least one of the $k$ generated postcondition sets is both test-correct and complete.

\noindent
\(\bullet\) {Correctness--completeness gap.}
To quantify the discrepancy between correctness and completeness, we define two complementary measures.
The \emph{absolute gap} is defined as
\begin{equation}
\Delta@k = \text{Corr@}k - \text{Comp@}k,
\end{equation}
which captures the extent to which test-based correctness overestimates completeness.
We also report the \emph{conditional completeness rate},
\begin{equation}
\rho@k = \frac{\text{Comp@}k}{\text{Corr@}k},
\end{equation}
measuring the fraction of test-correct postconditions that are also complete.
Unless otherwise specified, gap analyses in the main paper focus on \texttt{@1}, while multi-sample results are reported in the appendix.

\begin{table*}[t]

\setlength{\heavyrulewidth}{1.2pt}
\setlength{\lightrulewidth}{0.6pt}
\newcolumntype{Y}{>{\centering\arraybackslash}X}

\centering
\caption{
Postcondition generation evaluation results from Large Language Models.
The \textbf{bold} font emphasizes the highest-performing LLM for each metric within each language.
}
\label{tab:llm_results--app-2}

\tiny

\begin{adjustbox}{min width=\textwidth} % 或 max width=\linewidth

\begin{tabularx}{0.78\textwidth}{r|YY|YY|YY|YY|YY|YY}
\toprule

 & 
\multicolumn{4}{c|}{\textbf{C2P (Code-only)}} & 
\multicolumn{4}{c|}{\textbf{N2P (NL-only)}} & 
\multicolumn{4}{c}{\textbf{F2P (NL + Code)}} \\

\midrule

\multirow{2}{*}{\textbf{Metric}} 
& 
{\textbf{Corr}} & 
{\textbf{Comp}} & 
\multirow{2}{*}{\textbf{$\Delta$@1}} &
{\textbf{Comp/}} 
& 
{\textbf{Corr}} & 
{\textbf{Comp}} & 
\multirow{2}{*}{\textbf{$\Delta$@1}} &
{\textbf{Comp/}} 
& 
{\textbf{Corr}} & 
{\textbf{Comp}} & 
\multirow{2}{*}{\textbf{$\Delta$@1}} &
{\textbf{Comp/}} 
\\

 & \textbf{@1} & \textbf{@1} & & \textbf{Corr} 
 & \textbf{@1} & \textbf{@1} & & \textbf{Corr} 
 & \textbf{@1} & \textbf{@1} & & \textbf{Corr} 
 \\

\midrule

GPT-5 & 0.493 & \textbf{0.290} & 0.203 & 0.588 & 0.434 & \textbf{0.177} & 0.257 & 0.408 & 0.520 & \textbf{0.298} & 0.222 & 0.573 \\
Claude-4.5 & \textbf{0.645} & 0.234 & 0.411 & 0.362 & \textbf{0.593} & 0.150 & 0.443 & 0.253 & \textbf{0.650} & 0.237 & 0.413 & 0.364 \\
LLaMA-4 & 0.224 & 0.105 & 0.119 & 0.469 & 0.180 & 0.062 & 0.118 & 0.344 & 0.239 & 0.116 & 0.122 & 0.487 \\
Qwen3-32B & 0.141 & 0.062 & 0.079 & 0.441 & 0.092 & 0.043 & 0.049 & 0.469 & 0.120 & 0.060 & 0.061 & 0.494 \\
Gemma-3-27B & 0.117 & 0.054 & 0.063 & 0.459 & 0.097 & 0.026 & 0.071 & 0.266 & 0.117 & 0.053 & 0.064 & 0.453 \\

\bottomrule
\end{tabularx}

\end{adjustbox}

\end{table*}

\section{Correctness–Completeness Gap: Case study}
\label{sec:corr_vs_comp_cases}

\begin{figure*}[htbp] % h=here t=top b=bottom p=page of floats
  \centering
  \includegraphics[width=1\linewidth]{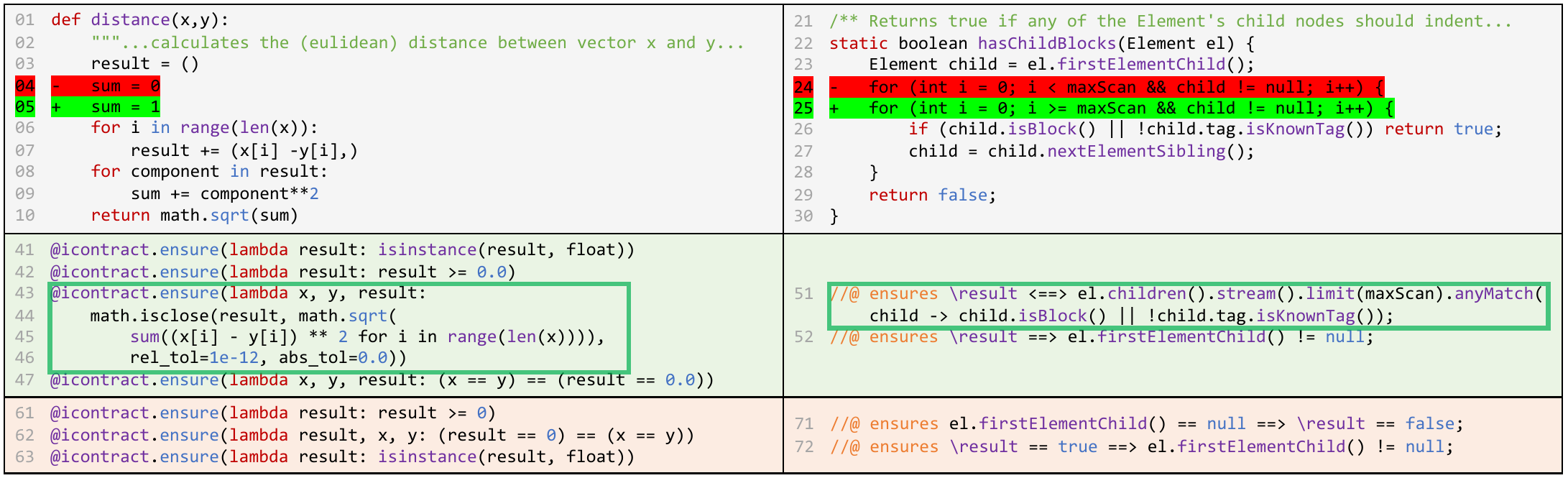}
  \caption{
  Left: A Python instance from \texttt{keon/algorithms} \cite{algorithms-keon-github}.
  Right: A Java instance from \texttt{jhy/jsoup} \cite{jsoup-jhy-github}.
  Top: Target methods. For each method, the ``+'' and ``-'' leading lines represent the differences from the original code to an example mutant.
  Middle: Complete postcondition sets (both are GPT-5 generated from the F2P task).
  Bottom: Correct but incomplete postcondition sets (Claude-4.5 generated from the F2P task).
  The green box enclosed the key postcondition that make the set more complete.
  }
  \label{fig:corr_vs_comp}
\end{figure*}

\cref{fig:corr_vs_comp} gives concrete examples.
For \texttt{distance} (left), the correct-but-incomplete set mainly checks boundary/special cases (e.g., \texttt{x==y}) and the return type, yet fails to kill a mutant that initializes \texttt{sum} to one when \texttt{x != y};
the complete set additionally checks the expected value (line~003), which rules out the mutant.
The Java example (right) shows a similar pattern: the correct-but-incomplete set includes a redundant constraint, while the complete set adds a stricter condition beyond the special case.

% \begin{takeawaybox}
% In summary, correctness does not necessarily imply high-quality postconditions: many correct postcondition sets are still incomplete.
% Completeness evaluation therefore provides a practical signal for assessing postcondition set quality beyond correctness.
% \end{takeawaybox}
\section{Ablation on Mutation Schemes} 
\label{sec:mutation_scheme_ablation}
% \juan{move to appendix}
\subsection{\camchg{False Discovery Rate Across Mutation Schemes}}

\begin{table*}[t]

\small

\setlength{\heavyrulewidth}{1.2pt}
\setlength{\lightrulewidth}{0.6pt}
\newcolumntype{Y}{>{\centering\arraybackslash}X}

\centering
\caption{
Mutation FDR (False Discovery Rate) ($\downarrow$) of different mutation schemes.
I.e., if we only use one scheme to identify the complete postconditions, the ratio of the discovered ``complete'' postcondition that cannot kill all mutants from another scheme.
}
\label{tab:mutation_precision}

\begin{tabularx}{1\textwidth}{r|YY|YY|YY|YY|YY|YY}
\toprule

 & 
\multicolumn{6}{c|}{\textbf{Python}} & 
\multicolumn{6}{c }{\textbf{Java}} \\

\midrule

 & 
\multicolumn{2}{c|}{\textbf{F2P}} & 
\multicolumn{2}{c|}{\textbf{C2P}} & 
\multicolumn{2}{c|}{\textbf{N2P}} & 
\multicolumn{2}{c|}{\textbf{F2P}} & 
\multicolumn{2}{c|}{\textbf{C2P}} & 
\multicolumn{2}{c}{\textbf{N2P}} 
\\

&
\textbf{\#Rule} & \textbf{\#LLM} & 
\textbf{\#Rule} & \textbf{\#LLM} & 
\textbf{\#Rule} & \textbf{\#LLM} & 
\textbf{\#Rule} & \textbf{\#LLM} &
\textbf{\#Rule} & \textbf{\#LLM} & 
\textbf{\#Rule} & \textbf{\#LLM} 
\\

\midrule

GPT-5 & 0.155 & \textbf{0.154} & \textbf{0.127} & 0.188 & \textbf{0.132} & 0.169 & 0.146 & \textbf{0.049} & 0.111 & \textbf{0.051} & 0.183 & \textbf{0.036} \\
Claude-4.5 & \textbf{0.164} & 0.201 & \textbf{0.153} & 0.205 & 0.209 & \textbf{0.152} & 0.220 & \textbf{0.064} & 0.210 & \textbf{0.049} & 0.250 & \textbf{0.061} \\
LLaMA-4 & 0.186 & \textbf{0.174} & \textbf{0.103} & 0.158 & \textbf{0.143} & 0.167 & 0.148 & \textbf{0.050} & 0.187 & \textbf{0.126} & 0.203 & \textbf{0.150} \\
Qwen3-32B & 0.400 & \textbf{0.000} & \textbf{0.000} & \textbf{0.000} & - & - & 0.181 & \textbf{0.144} & 0.227 & \textbf{0.077} & 0.108 & \textbf{0.042} \\
Gemma-3-27B & 0.667 & \textbf{0.500} & 0.833 & \textbf{0.000} & - & - & 0.252 & \textbf{0.097} & 0.215 & \textbf{0.167} & 0.270 & \textbf{0.062} \\
\midrule
\textbf{Avg} & 0.314 & \textbf{0.206} & 0.243 & \textbf{0.110} & \textbf{0.161} & 0.162 & 0.189 & \textbf{0.081} & 0.190 & \textbf{0.094} & 0.203 & \textbf{0.070} \\

\bottomrule
\end{tabularx}
\end{table*}

We ablate our two mutation schemes---rule-based and LLM-based---to assess their complementarity.
\cref{tab:mutation_precision} reports the false discovery rate (FDR) for each scheme.
Concretely, for a given scheme $S$, we first evaluate completeness using only mutants generated by $S$~and collect postcondition sets that are classified as \emph{complete}~under this restricted mutant set.
We then re-evaluate these same postcondition sets on mutants from the other scheme $\bar{S}$~and compute the fraction that fail to kill all $\bar{S}$-mutants; this fraction is the FDR.
A higher FDR indicates that a single scheme is insufficient in isolation, and that the other scheme contributes additional mutants that expose incompleteness.

As shown in \cref{tab:mutation_precision}, both schemes exhibit non-trivial FDRs, suggesting meaningful complementarity.
For example, the Avg row shows an FDR of 0.206 for LLM-based mutation on Python--F2P, and an FDR of 0.203 for rule-based mutation on Java--N2P.
In other words, if we rely on only one mutation scheme, a noticeable fraction of postcondition sets deemed ``complete'' would still be incomplete and would be revealed by mutants from the other scheme.

Overall, rule- and LLM-based mutation complement each other for completeness evaluation.

\subsection{\camchg{Complementarity Under Scheme Exclusion}}

\begin{table*}[t]

\small

\setlength{\heavyrulewidth}{1.2pt}
\setlength{\lightrulewidth}{0.6pt}
\newcolumntype{Y}{>{\centering\arraybackslash}X}

\centering
\caption{
\camcaption{Comp@1 after excluding one mutation family from completeness evaluation.}
\camcaption{Higher values after exclusion indicate that the removed mutation family exposes defects not covered well by the retained family.}
}
\label{tab:mutation_complementarity}

\begin{tabularx}{\textwidth}{l|YYYYY}
\toprule

\textbf{Condition} & \textbf{GPT-5} & \textbf{Claude-4.5} & \textbf{LLaMA-4} & \textbf{Qwen3-32B} & \textbf{Gemma-3-27B} \\

\midrule

w/ all mutants & 0.2551 & 0.2068 & 0.0944 & 0.0551 & 0.0441 \\
wo/ LLM-based & 0.3094 & 0.2944 & 0.1046 & 0.0606 & 0.0481 \\
wo/ operator-based & 0.2773 & 0.2289 & 0.1281 & 0.0690 & 0.0622 \\

\bottomrule
\end{tabularx}
\end{table*}

\begin{camchgblock}
We further recompute Comp@1 after excluding either all LLM-based mutants or all operator-based mutants.
\Cref{tab:mutation_complementarity} shows that scores increase after removing either mutation family, indicating that each family exposes defects the other does not cover well.
For example, when LLM-based mutants are excluded, Claude-4.5 increases from 0.2068 to 0.2944; when operator-based mutants are excluded, LLaMA-4 increases from 0.0944 to 0.1281.
This provides a direct complement to the FDR analysis above: high completeness requires handling both mutation families well.
\end{camchgblock}

\subsection{\camchg{Mutant-Budget Stability}}

\begin{table*}[t]

\small

\setlength{\heavyrulewidth}{1.2pt}
\setlength{\lightrulewidth}{0.6pt}
\newcolumntype{Y}{>{\centering\arraybackslash}X}

\centering
\caption{
\camcaption{Comp@1 under two forms of large mutant-set perturbation.}
\camcaption{Except for the baseline column, each cell reports the mean and the standard deviation over 50 trials.}
}
\label{tab:mutation_stability}

\centering
\begin{adjustbox}{max width=\textwidth}
\begin{tabular}{l|c|cc|ccccc}
\toprule

\multirow{2}{*}{\textbf{Model}} &
\multirow{2}{*}{\textbf{Baseline}} &
\multirow{2}{*}{\shortstack{\textbf{Ex. 5}\\\textbf{Operators}}} &
\multirow{2}{*}{\shortstack{\textbf{Ex. 50\%}\\\textbf{LLM Mutants}}} &
\multicolumn{5}{c}{\textbf{Mutant-Budget Fractions}} \\

& & & &
\textbf{0.1} & \textbf{0.2} & \textbf{0.8} & \textbf{0.9} & \textbf{1.0} \\

\midrule

\multirow{2}{*}{GPT-5} & \multirow{2}{*}{0.255} & 0.265 & 0.265 & 0.331 & 0.307 & 0.262 & 0.259 & 0.255 \\
 &  & (0.004) & (0.002) & (0.007) & (0.006) & (0.003) & (0.003) & (0.000) \\
\midrule
\multirow{2}{*}{Claude-4.5} & \multirow{2}{*}{0.207} & 0.215 & 0.216 & 0.298 & 0.261 & 0.212 & 0.209 & 0.207 \\
 &  & (0.003) & (0.002) & (0.010) & (0.008) & (0.003) & (0.002) & (0.000) \\
\midrule
\multirow{2}{*}{LLaMA-4} & \multirow{2}{*}{0.094} & 0.098 & 0.098 & 0.129 & 0.116 & 0.097 & 0.096 & 0.094 \\
 &  & (0.002) & (0.001) & (0.006) & (0.004) & (0.002) & (0.002) & (0.000) \\
\midrule
\multirow{2}{*}{Qwen3-32B} & \multirow{2}{*}{0.055} & 0.056 & 0.057 & 0.074 & 0.068 & 0.056 & 0.056 & 0.055 \\
 &  & (0.001) & (0.001) & (0.003) & (0.003) & (0.001) & (0.001) & (0.000) \\
\midrule
\multirow{2}{*}{Gemma-3-27B} & \multirow{2}{*}{0.044} & 0.046 & 0.045 & 0.059 & 0.053 & 0.045 & 0.045 & 0.044 \\
 &  & (0.001) & (0.001) & (0.004) & (0.003) & (0.001) & (0.001) & (0.000) \\

\bottomrule
\end{tabular}
\end{adjustbox}
\end{table*}

\begin{camchgblock}
We also vary the overall mutant budget by randomly sampling a fixed fraction of the full mutant set for each method.
As shown in \Cref{tab:mutation_stability}, completeness approaches saturation as the budget increases.
Between 80\%, 90\%, and 100\% of mutants, the mean Comp@1 changes only slightly for all five models, and the standard deviations are small once the budget reaches 80\% or above.
This suggests that completeness is not dominated by a small number of unusually influential mutants.
\end{camchgblock}

\subsection{\camchg{Operator-Level Ablation}}

\begin{table}[t]

\footnotesize

\setlength{\heavyrulewidth}{1.2pt}
\setlength{\lightrulewidth}{0.6pt}

\centering
\caption{
\camcaption{Operator-level ablation for rule-based mutation, averaged across models.}
}
\label{tab:operator_ablation}

\begin{tabularx}{\columnwidth}{>{\raggedright\arraybackslash}p{0.66\columnwidth}r}
\toprule

\multicolumn{2}{c}{\textbf{Python}} \\
\textbf{Excluded operator} & \textbf{Comp@1} \\

\midrule

No exclusion (baseline) & 0.0516 \\
asymmetric\_str\_method\_swap & 0.0516 \\
augassign\_to\_assign & 0.0516 \\
numeric\_increments & 0.0516 \\
symmetric\_str\_method\_swap & 0.0516 \\
unary\_op\_removal & 0.0516 \\
boolean\_constant\_flip & 0.0519 \\
arg\_removal & 0.0521 \\
string\_perturbation & 0.0523 \\
keyword\_rewrite & 0.0525 \\
operator\_replacement & 0.0531 \\
assignment\_nullification & 0.0554 \\

\midrule
\multicolumn{2}{c}{\textbf{Java}} \\
\textbf{Excluded operator} & \textbf{Comp@1} \\

\midrule

No exclusion (baseline) & 0.2106 \\
conditionals\_boundary & 0.2106 \\
false\_returns & 0.2106 \\
invert\_negatives & 0.2106 \\
true\_returns & 0.2107 \\
increments & 0.2108 \\
math & 0.2109 \\
primitive\_returns & 0.2110 \\
negate\_conditionals & 0.2110 \\
empty\_returns & 0.2111 \\
void\_method\_call & 0.2152 \\
null\_returns & 0.2169 \\

\bottomrule
\end{tabularx}
\end{table}

\begin{camchgblock}
To test whether the metric is overly sensitive to a particular rule-based operator, we remove each operator in turn and average the resulting Comp@1 across models.
\Cref{tab:operator_ablation} shows that most removals have only minor effects.
In Python, dropping any single operator changes Comp@1 by at most 0.0038; in Java, the largest change is 0.0063 after removing \texttt{null\_returns}.
This supports that completeness is not driven by one critical operator.
\end{camchgblock}

\subsection{\camchg{Mutation-Strength Ablation}}

\begin{camchgblock}
Finally, we test robustness under large random perturbations of the mutant set.
We either remove 5 of the 11 rule-based operators at random or remove 50\% of LLM-based mutants, and then recompute Comp@1 over 50 trials.
As shown in \Cref{tab:mutation_stability}, the mean changes are small and the standard deviations remain low for all models.
These results indicate that the completeness metric is stable under substantial changes in both mutant composition and mutant strength.
\end{camchgblock}

% \subsection{Ablation on Mutation Schemes}

% For ablation, we analyze the complementarity between the two mutation schemes we used, i.e., rule-based mutation and LLM-based mutation.
% The \cref{tab:mutation_precision}~demonstrates the FDR (False Discovery Rate) of different mutation schemes. 
% In this experiment, we only keep one mutation scheme and exclude the other for a completeness evaluation. 
% Then we demonstrate the ratio of these ``partial-identified-complete'' postconditions that cannot kill all mutants from another mutation (i.e., false discovery rate).
% A higher false discovery rate indicates that the mutants are less capable of evaluating completeness when generated by a single mutation scheme, i.e., a higher complementarity level between different mutation schemes.
% From \cref{tab:mutation_precision}, we can observe that a relatively high FDR exists for both rule-based and LLM-based mutations.
% %
% For example, the LLM-based mutation achieves a 0.206 average FDR in the Python-F2P tasks, and the rule-based mutation FDR is 0.203 in the Java-N2P tasks.
% In other words, if we only use one mutation scheme, a relatively high ratio of identified complete post-conditions would actually be incomplete and could be correctly identified by other mutations.

% In conclusion, the mutation schemes we used are complementary to each other.

\begin{figure*}[htbp] % h=here t=top b=bottom p=page of floats
  \centering
  \includegraphics[width=1\linewidth]{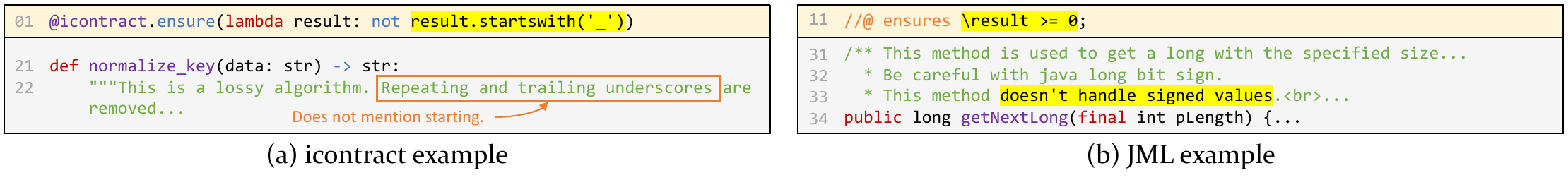}
  \caption{
  Two examples of incorrect postconditions generated due to LLMs' hallucination on semantics.
  The left side is the generated postconditions from the N2P task and the target method from \texttt{kellyjonbrazil/jc}.
  The documentation states that repeating and trailing underscores
  are removed, but makes no claim about leading underscores.
  The generated postcondition nevertheless enforces the absence of leading underscores,
  introducing an unsupported constraint.
  The right side target method is from \texttt{devnied/Bit-lib4j}. The postcondition is generated from the F2P task.
  The method reads a \texttt{long} value of a given bit length.
  Although the comment notes that signed values are ``not handled''
  this implies no special treatment of the sign bit rather than non-negativity.
  The generated postcondition incorrectly enforces \texttt{\textbackslash result >= 0},
  thereby ruling out valid behaviors.
  Both postconditions are generated by Claude-4.5.
  }
  \label{fig:hall_semantics}
\end{figure*}

\begin{figure*}[t] % h=here t=top b=bottom p=page of floats
  \centering
  \includegraphics[width=1\linewidth]{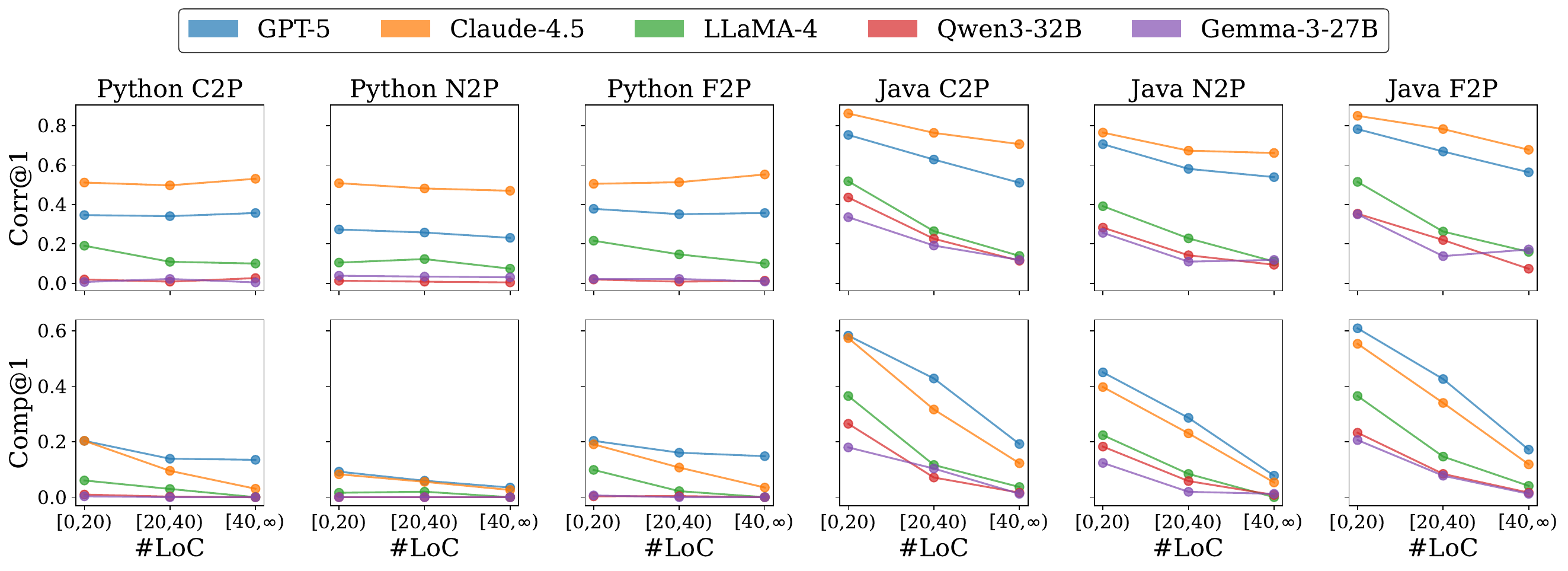}
  \caption{
Corr@1 (top) and Comp@1 (bottom) across target methods grouped by lines of code (LoC).
}
  \label{fig:lines}
\end{figure*}
\begin{table*}[t]
\small
\setlength{\heavyrulewidth}{1.2pt}
\setlength{\lightrulewidth}{0.6pt}
\newcolumntype{Y}{>{\centering\arraybackslash}X}

\centering
\caption{
Postcondition generation performance on non-standalone (\textbf{Dep.}) and standalone (\textbf{Solo.}) methods.
\textbf{Dep.} requires repository-specific dependencies, while \textbf{Solo.} is runnable with only built-ins.
Bold indicates the higher score within each comparison.
}\label{tab:dependency_table_JZ}

\begin{tabularx}{0.83\textwidth}{rr|YY|YY|YY|YY}
\toprule
\multicolumn{2}{r|}{\textbf{Language}} &
\multicolumn{4}{c|}{\textbf{Python}} &
\multicolumn{4}{c}{\textbf{Java}} \\
\midrule
\multicolumn{2}{r|}{\multirow{2}{*}{\textbf{Model / Setting}}} &
\multicolumn{2}{c|}{\textbf{Corr@1}} &
\multicolumn{2}{c|}{\textbf{Comp@1}} &
\multicolumn{2}{c|}{\textbf{Corr@1}} &
\multicolumn{2}{c}{\textbf{Comp@1}} \\
& &
\textbf{Dep.} & \textbf{Solo.} &
\textbf{Dep.} & \textbf{Solo.} &
\textbf{Dep.} & \textbf{Solo.} &
\textbf{Dep.} & \textbf{Solo.} \\
\midrule

\multirow{3}{*}{GPT-5}
 & C2P & 0.315 & \textbf{0.419} & 0.135 & \textbf{0.210} & 0.614 & \textbf{0.768} & 0.385 & \textbf{0.600} \\
 & N2P & 0.211 & \textbf{0.365} & 0.038 & \textbf{0.126} & 0.591 & \textbf{0.708} & 0.245 & \textbf{0.503} \\
 & F2P & 0.335 & \textbf{0.419} & 0.139 & \textbf{0.245} & 0.662 & \textbf{0.768} & 0.393 & \textbf{0.578} \\
\addlinespace[2pt]\cmidrule(lr){1-2}

\multirow{3}{*}{Claude-4.5}
 & C2P & 0.470 & \textbf{0.600} & 0.070 & \textbf{0.216} & 0.782 & \textbf{0.784} & 0.332 & \textbf{0.459} \\
 & N2P & 0.441 & \textbf{0.597} & 0.018 & \textbf{0.152} & 0.699 & \textbf{0.703} & 0.216 & \textbf{0.368} \\
 & F2P & 0.484 & \textbf{0.603} & 0.072 & \textbf{0.223} & 0.775 & \textbf{0.805} & 0.335 & \textbf{0.459} \\
\addlinespace[2pt]\cmidrule(lr){1-2}

\multirow{3}{*}{LLaMA-4}
 & C2P & 0.120 & \textbf{0.158} & 0.024 & \textbf{0.052} & 0.301 & \textbf{0.395} & 0.160 & \textbf{0.265} \\
 & N2P & 0.092 & \textbf{0.142} & 0.004 & \textbf{0.039} & 0.232 & \textbf{0.351} & 0.083 & \textbf{0.232} \\
 & F2P & 0.143 & \textbf{0.190} & 0.023 & \textbf{0.081} & 0.302 & \textbf{0.405} & 0.173 & \textbf{0.281} \\
\addlinespace[2pt]\cmidrule(lr){1-2}

\multirow{3}{*}{Qwen3-32B}
 & C2P & 0.012 & \textbf{0.023} & 0.001 & \textbf{0.010} & 0.250 & \textbf{0.351} & 0.116 & \textbf{0.146} \\
 & N2P & \textbf{0.009} & 0.006 & 0.000 & 0.000 & 0.149 & \textbf{0.303} & 0.055 & \textbf{0.232} \\
 & F2P & \textbf{0.015} & 0.006 & 0.003 & \textbf{0.003} & 0.215 & \textbf{0.292} & 0.108 & \textbf{0.157} \\
\addlinespace[2pt]\cmidrule(lr){1-2}

\multirow{3}{*}{Gemma-3-27B}
 & C2P & \textbf{0.016} & 0.006 & \textbf{0.001} & 0.000 & 0.209 & \textbf{0.276} & 0.090 & \textbf{0.184} \\
 & N2P & \textbf{0.035} & 0.032 & 0.000 & 0.000 & \textbf{0.170} & 0.108 & 0.046 & \textbf{0.076} \\
 & F2P & 0.019 & \textbf{0.019} & \textbf{0.003} & 0.000 & 0.197 & \textbf{0.297} & 0.083 & \textbf{0.200} \\
\midrule

\multicolumn{2}{c|}{\textbf{Avg}} &
0.181 & \textbf{0.239} &
0.035 & \textbf{0.090} &
0.410 & \textbf{0.488} &
0.188 & \textbf{0.316} \\
\bottomrule
\end{tabularx}
\end{table*}

\begin{table}[t]

\small

\setlength{\heavyrulewidth}{1.2pt}
\setlength{\lightrulewidth}{0.6pt}
\newcolumntype{Y}{>{\centering\arraybackslash}X}

\centering
\caption{
\camcaption{Effect of adding grammar guidance on the F2P task for two strong models.}
}
\label{tab:grammar_guidance_f2p}

\begin{tabularx}{\columnwidth}{l|YY|YY}
\toprule

\multirow{2}{*}{\textbf{Model}} & \multicolumn{2}{c|}{\textbf{w/ guidance}} & \multicolumn{2}{c}{\textbf{w/o guidance}} \\
& \textbf{Corr@1} & \textbf{Comp@1} & \textbf{Corr@1} & \textbf{Comp@1} \\

\midrule

GPT-5 & 0.767 & 0.400 & 0.520 & 0.298 \\
Claude-4.5 & 0.814 & 0.283 & 0.629 & 0.207 \\

\bottomrule
\end{tabularx}
\end{table}

\begin{figure}[htbp] % h=here t=top b=bottom p=page of floats
  \centering
  \includegraphics[width=1\linewidth]{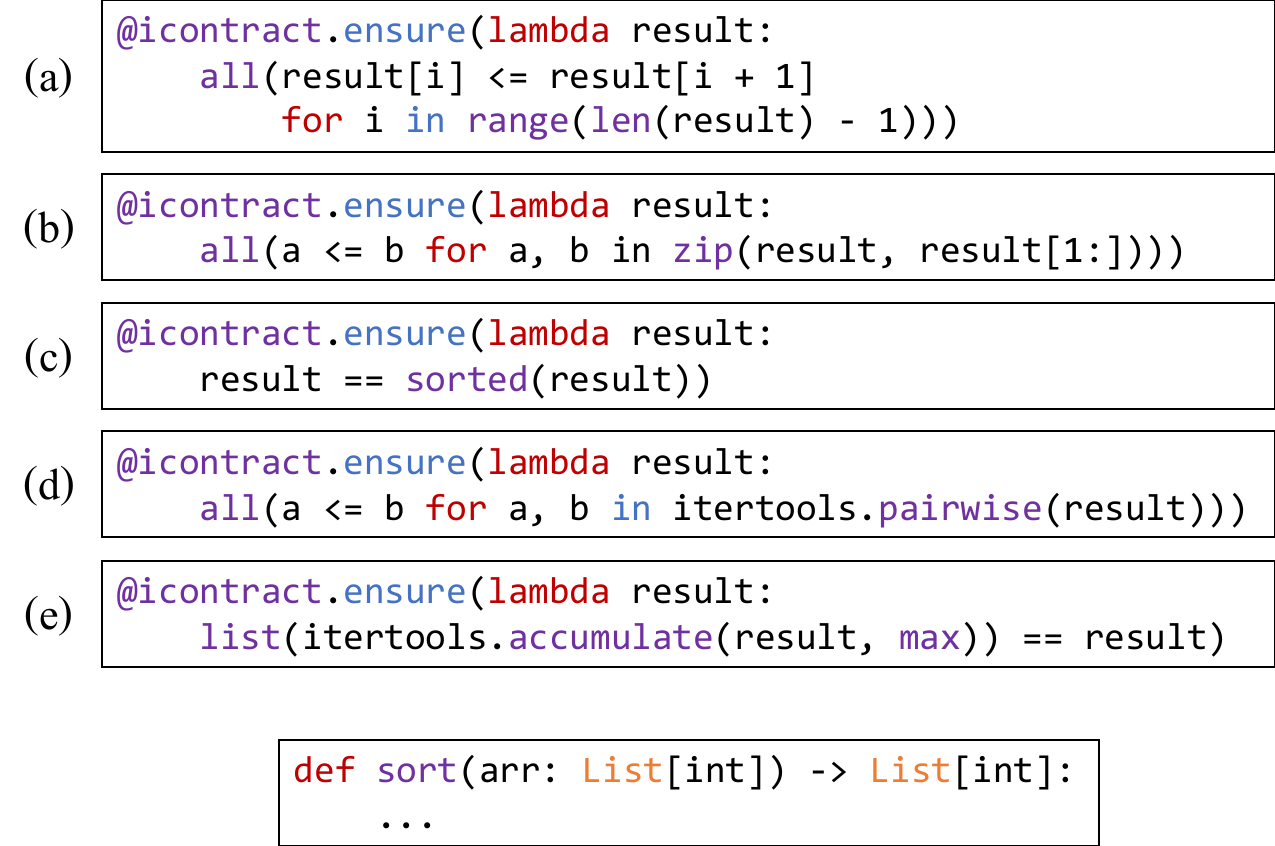}
  \caption{
  An example for sorting algorithm validation.
  (a) to (e) are postconditions written in icontract, which are semantically equivalent but syntactically different. 
  }
  \label{fig:sorting}
\end{figure}

\begin{figure}[t] % h=here t=top b=bottom p=page of floats
  \centering
  \includegraphics[width=1\linewidth]{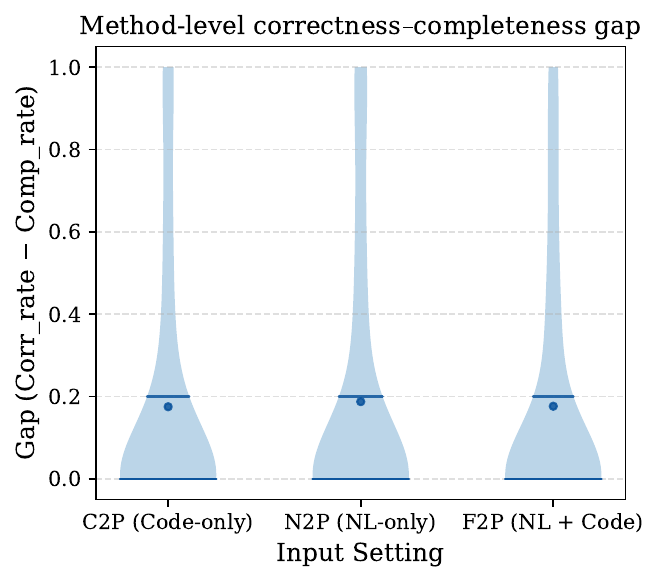}
  \caption{
  Distribution of method-level correctness–completeness gaps under different input settings, without methods with Corr@k = 0 excluded.
  The gaps are relatively smaller than \cref{fig:violin--main}~due to generally low Corr\_rates (e.g., Corr@1 = 0.224 in \cref{tab:llm_results--app-2}~for LLaMA-4 under C2P setting)
  }
  \label{fig:violin--main--unfilter}
\end{figure}

\newenvironment{codefigure}[2]{%
\def\codefigurecaption{#1}%
\def\codefigurelabel{#2}%
\begin{figure*}[p]
\centering
\begin{tcolorbox}[width=0.96\textwidth,colback=white,colframe=black,boxrule=0.5pt,arc=0pt,left=6pt,right=6pt,top=6pt,bottom=6pt]
\footnotesize
}{%
\end{tcolorbox}
\caption{\camcaption{\codefigurecaption}}
\label{\codefigurelabel}
\end{figure*}
}

\section{\camchg{Discussion on Bias from LLM-Generated Ground Truths}}
\begin{camchgblock}
As described in Section \ref{postcond_generation}, we introduce an LLM (GPT-5-mini) for ground truth generation. 66\% of the final benchmark ground-truths are LLM-generated.
In this section, we discuss whether the LLM-generated ground truths will introduce bias in the benchmark evaluation.

Importantly, \projName~does not use ground-truth postconditions during evaluation. Corr@1 and Comp@1 are computed solely via test execution and mutant discrimination, independent of any reference specification. Therefore, adopting LLM-generated outputs as ground truth does not affect how models are scored.
During dataset construction, LLM was used only to reduce manual effort by proposing candidate ground-truths. Target methods are filtered solely when human experts judge that writing a complete postcondition set is infeasible under the current framework (\autoref{sec:expert_refinement}). LLM assists in identifying solvable cases but does not influence evaluation criteria or introduce model-specific bias.
\end{camchgblock}

\section{Expert Refinement Statistics/Analysis}
\label{sec:expert_refinement}
As Section \ref{expert_refinement}~mentioned, during domain expert refinement, some methods/mutants are found inappropriate for inclusion in the benchmark. 
We analyze and attribute them in this section.

\begin{camchgblock}
Overall, filtering affects a minority of the sampled methods rather than most of them.
Among 297 sampled Python methods, 87 are filtered (29.3\%); among 234 sampled Java methods, 24 are filtered (10.3\%).
Most filtered methods arise from limitations of the current specification frameworks rather than the complexity of the methods: 45 of the 87 filtered Python methods and 13 of the 24 filtered Java methods fall into framework-limited categories, such as iterator-related or concurrency-related.
For the remaining filtered cases, domain experts determined that writing a complete postcondition set would still be non-trivial even for humans under the current evaluation setup.
\end{camchgblock}

\subsection{Python Language}
297 Python methods sampled by farthest-first traversal. We ensure that the sampled methods have at least five mutants.

Among them, 135 target methods have complete postconditions generated by GPT-5-mini. These methods have 2670 mutants, all of which are killed by auto-generated postconditions.

We manually checked the remaining 162 Python methods and attempted to write complete postconditions for them.

Of the 162 Python methods, 75 have manually written postconditions that are identified as complete. The remaining 87 methods are identified as inappropriate for inclusion. The reasons for which they are filtered are analyzed in the following section.

These 75 methods have 2957 mutants. Among them, 43 mutants were identified as inappropriate for inclusion and were filtered. 10 methods filtered the mutants. The reasons for their filtration are analyzed in the following section.

Overall, we have 135 + 75 = 210 Python methods in our benchmark.

\subsubsection{Method-level Filtering Analysis}

After manual checking, 87 methods are filtered. Under the current formal specification framework, writing complete postcondition expressions for them is non-trivial even for human engineers. We analyze the reason for filtering.

\paragraph{Iterator related (18).}
\camchg{By limitation of the API, Iterator/Generator objects can only be iterated over once. Validating them by postcondition expressions has side effects. Therefore, methods have key variables that in Iterator/Generator types, and the key feature relies on accessing elements from the iterator objects that have been filtered. \Cref{fig:python-filter-iterator-prefix-lines} shows the \texttt{prefix\_lines} method from project miyuchina/mistletoe.}

Validating these methods requires access to the actual string elements. However, both the input parameters and return value have Iterator objects. Accessing elements in \texttt{@icontract.snapshot}~or \texttt{@icontract.ensure}~decorators leads to side-effects. The current icontract does not provide an effective path to deal with these cases.

\paragraph{Timeout (17).}
The postcondition running also requires time. 
\camchg{For these cases, complete validations lead to timeouts. \Cref{fig:python-filter-timeout-load-inventory} shows the \texttt{load\_inventory}~method from the project hynek/doc2dash.}
Note that the return value is directly related to the content in the \texttt{objects.inv}~file. 
Therefore, a complete validation requires accessing the file content, leading to timeouts.

\paragraph{Concurrency related (13).}
Some methods' key feature is related to obtaining results through concurrency mechanisms (starting sub-processes, sub-threads, etc.), which are identified as inappropriate for inclusion.
\camchg{\Cref{fig:python-filter-concurrency-execute-code} shows method \texttt{execute\_code} in vndee/llm-sandbox.}
In this case, the results depend on running the inputted \texttt{code}~inside the session, which launches a sandbox using Docker, Kubernetes, Podman, etc.
This leads to a non-trivial validation, since it would require analyzing the code's semantics or launching a new sandbox within the postcondition.

\camchg{Moreover, this category also includes several coroutine-using methods. \Cref{fig:python-filter-concurrency-get-task} shows \texttt{get\_task}~in a2aproject/a2a-python.}
Since the return value directly relies on the results from \texttt{self.task\_store.get}.
If \texttt{get}~is a sync method, we can call it and validate on its return value inside the postcondition.
However, since \texttt{get}~is an async method and Python does not support async lambda. With icontract, defining async conditions (and snapshots) is indeed tedious \cite{icontract-doc-async}.

\paragraph{Language entities related (8).}
\camchg{Some methods return a function object, edit a class object, etc. Validation on these language entities is generally non-trivial. \Cref{fig:python-filter-language-while-until-true} shows \texttt{while\_until\_true}~in pypyr/pypyr, which returns a function object.}
\camchg{Moreover, \Cref{fig:python-filter-language-make-sync} shows \texttt{make\_sync}~in run-llama/llama\_deploy, which edits and returns a class object.}

\paragraph{Ordinary (9).}
Some methods' key features (logging, printing to stdout, sending email, etc.) are not related to changes in the accessible code context status and cannot be caught by the specification framework.
\camchg{\Cref{fig:python-filter-ordinary-rich-format-error} shows \texttt{rich\_format\_error} in fastapi/typer.}
This method's key feature is to print error logs under different conditions. Buggy mutants that disturb log printing can be caught by the test suite since the test cases use mock util to validate logging outputs. However, this method is not feasible to validate with only the formal specification framework. 

\paragraph{Specified error process (5).}
For Python, we need the program output \texttt{icontract.errors.ViolationError}~in stderr to know that the postcondition can kill the mutant.
However, in some cases, errors do not output an error log. 
\camchg{\Cref{fig:python-filter-specified-error-help-create} shows the test case format for \texttt{\_make\_command\_help}~in langchain-ai/langchain-mcp-adapters.}
These test cases would simply fail due to assertion error (\texttt{exit\_code} is not 0), and the error message will not specify whether the error is from icontract violation error.

\paragraph{Others (17).}
Not included above, and non-trivial for human engineers to write complete postconditions.
These methods are generally lengthy or exist within complex code contexts, increasing the costs of understanding. 
The human engineer failed to write complete postconditions for these. 
\camchg{\Cref{fig:python-filter-others-merge} shows method \texttt{merge} in pypyr/pypyr.}
This 81-line method is complicated, and recursion is used. The human engineer failed to write complete postconditions for it.

\subsubsection{Mutant-level Filtering Analysis}
Of the 210 Python methods in the benchmark, 10 have mutants filtered after manual checking. 43 mutants are filtered. Note that we ensure the main feature of the methods can still be validated after some of the mutants are filtered for these methods.

% Preamble (if not already):
% \usepackage{fancyvrb}

\paragraph{Ordinary (29 mutants; five methods involved).}
The feature that mutant disturbs (logging, process sleeping length, etc.) is not related to accessible code context status change and cannot be caught by the specification framework.
\camchg{\Cref{fig:python-mutant-ordinary-getprops} shows \texttt{getprops} in \texttt{google/mobly}.}

The main feature of this method is validatable. However, a few mutants disturb the line of \texttt{time.sleep}, e.g., replacing it by \texttt{time.sleep(0)}.
We filter out these mutants for a similar reason to the way we filter ``ordinary'' category methods.

\paragraph{Specified error process (nine mutants; two methods involved).}
These mutants triggered errors that do not output error types.
\camchg{\Cref{fig:python-mutant-specified-error-parse-ignore} shows the triggered test case format for two mutants on method \texttt{\_parse\_ignore\_file} in project \texttt{coderamp-labs/gitingest}.}

These test cases would simply fail due to assertion error (\texttt{exit\_code} is not 0), and the error message will not specify whether the error is from \texttt{icontract} violation error.
For this case, since only a few mutants (seven of 34) are affected by this issue, we keep the method but filter out affected mutants.

\paragraph{Equivalent but not the same results (two mutants; one method involved).}
Sometimes, the algorithm has multiple possible results, the test cases are over-validated, and fail on mutants that are actually equivalent to the correct code.
We filter out these mutants since we only keep buggy mutants.
\camchg{\Cref{fig:python-mutant-equivalent-k-closest} shows the original method \texttt{k\_closest(points, k, origin)} from \texttt{keon/algorithms}, the equivalent mutant, and the over-constrained test case.}

While the mutant's return value \texttt{[(-2, -2), (-1, 0), (1, 1), (1, 0)]} is equivalent to the expected result, the test case failed.

Therefore, we filter out mutants under this category since we only keep buggy mutants.

\paragraph{Timeout (three mutants; two methods involved).}
The postcondition running also requires time. Post-conditions that are complete enough to kill these mutants lead to a timeout.
\camchg{\Cref{fig:python-mutant-timeout-partition-by-spaces} shows the \texttt{partition\_by\_spaces} method in the \texttt{frostming/marko} project together with the downstream call path that amplifies the failure into a timeout.}

However, one mutant replaces the \texttt{break} in \texttt{partition\_by\_spaces} with \texttt{return}, directly returning nothing.
That will cause an error in \texttt{break\_paragraph}. Further, it leads to more loop running in the \texttt{parse} method, and finally causes a timeout.
Since only one mutant triggers this timeout issue, and the main feature of \texttt{partition\_by\_spaces} is still validatable, we keep this target method and filter out this mutant.

\subsection{Java Language}
We sampled 234 Java methods by farthest-first traversal. We ensure that the sampled methods have at least five mutants.

Among them, 141 target methods have complete postconditions generated by GPT-5-mini. These methods have 2{,}530 mutants, all of which are killed by auto-generated postconditions.

We manually checked the remaining 93 Java methods, trying to write complete postconditions for them.

From the 93 Java methods, 69 methods have manually-written postconditions that are identified as complete. The remaining 24 methods are identified as inappropriate for inclusion. The reasons that they are filtered are analyzed in the following section.

These 69 methods have 2{,}075 mutants. Among them, two mutants were identified as inappropriate to be included and filtered. One method has mutants being filtered. The reasons that they are filtered are analyzed in the following section.

Overall, we have $141 + 69 = 210$ Java methods in our benchmark.

\subsubsection{Method-level Filtering Analysis}

After manual checking, 24 methods are filtered. Under the current formal specification framework, writing complete postcondition expressions for them is non-trivial even for human engineers. We analyze the reason for filtering.

\paragraph{Timeout (5).}
\camchg{The postcondition running also requires time. For these cases, complete validations lead to timeouts. \Cref{fig:java-filter-timeout-compare} shows the \texttt{compare} method from \texttt{red6/pdfcompare}.}

This method applies a PDF loading and comparison. Therefore, validating this method generally requires accessing two PDF files, leading to timeouts.

\paragraph{Iterator related (4).}
\camchg{By limitation of the API, Iterator/Generator objects can only be iterated over once. Validating them by postcondition expressions has side effects. Therefore, methods have key variables that in Iterator/Generator types, and the key feature relies on accessing elements from the iterator objects that have been filtered. \Cref{fig:java-filter-iterator-eo-yaml} shows the \texttt{iterator} method from project \texttt{decorators-squad/eo-yaml}.}

Validating the return value of this method requires access to the actual string elements of the return value. However, the return value has Iterator objects. Accessing elements in postconditions leads to side-effects.

\paragraph{Inaccessible code context (4).}
\camchg{In Java, only public cross-class fields/methods can be accessed. This category refers to the case where the target method modifies some write-only code context. The postcondition cannot perform validation due to inaccessible code contexts. \Cref{fig:java-filter-inaccessible-draw-rectangles} shows \texttt{drawRectanglesOfDifferences} in \texttt{romankh3/image-comparison}.}

The \texttt{java.awt.Graphics2D} object is created from \texttt{java.awt.image.BufferedImage} by \texttt{BufferedImage::createGraphics()} API. However, due to the privacy limitation, the buffered image cannot be accessed by the created \texttt{Graphics2D} object. This method \texttt{drawRectanglesOfDifferences} draws rectangles onto a \texttt{BufferedImage} object using a \texttt{Graphics2D} object \texttt{graphics}, while the modified buffered image cannot be accessed for validation.

\paragraph{Random related (3).}
\camchg{Some methods rely on random algorithms. For defects within these methods, test cases can be called multiple times, and bugs can be distinguished by observing the statistics, while a postcondition that runs each time after the method calls cannot catch the bugs. \Cref{fig:java-filter-random-change-mess} shows \texttt{changeMess} in \texttt{TheAlgorithms/Java}.
% , and \Cref{fig:java-filter-random-change-mess-test} shows a representative test case.
}

\camchg{\texttt{ber} is a field. This method randomly gets a pseudorandom \texttt{x} between 0 and 1 and modifies \texttt{message} when \texttt{x < ber}. 
\Cref{fig:java-filter-random-change-mess-test} shows one representative test case that covers this method.
}
After 1{,}000 calls of \texttt{changeMess}, it is nearly certain that the \texttt{wrongMess} field is larger than zero. However, the postcondition validation is called each time after the method runs. It is possible that after a method call, the \texttt{wrongMess} is still zero. Therefore, it is non-trivial to validate \texttt{wrongMess} due to this randomness.

\paragraph{Ordinary (2).}
\camchg{Some methods' key feature is not related to accessible code context status change and cannot be caught by the specification framework. \Cref{fig:java-filter-ordinary-execute} shows \texttt{execute} in \texttt{antkorwin/xsync}.}

As the key feature of this \texttt{execute} method, we need to validate whether the \texttt{runnable} has been run, which is non-trivial to validate since the implementation of \texttt{runnable} can be various.

\paragraph{Others (6).}
\camchg{Not included above, and non-trivial for human engineers to write complete postconditions. These methods are generally long or within complex code contexts, increasing the understanding costs. \Cref{fig:java-filter-others-douglas-peucker} shows \texttt{douglasPeucker} in \texttt{dyn4j/dyn4j}.}

This 85-line method is complicated, and recursion is used. The human engineer failed to write complete postconditions for it.

\subsubsection{Mutant-level Filtering Analysis}

Of the 210 Java methods in the benchmark, one has mutants filtered after manual checking. Two mutants are filtered. Note that we ensure the main feature of the methods can still be validated after part of the mutants are filtered for these three methods.

All mutants are filtered due to \textbf{nameless implementation}, which happens in method \texttt{accumulate} from \texttt{dyn4j/dyn4j}:
\camchg{\Cref{fig:java-mutant-nameless-accumulate} shows the target method, and \Cref{fig:java-mutant-nameless-accumulate-test} shows a representative \texttt{isComplete} implementation from the test suite.}

This method has two mutants that replace \texttt{if (force.isComplete(elapsedTime))} and \texttt{if (torque.isComplete(elapsedTime))} with \texttt{if (elapsedTime > 0)}. If \texttt{isComplete} has no side-effect, we can call it in the postconditions and validate that after the method calls, no element in list \texttt{this.forces} and \texttt{this.torques} satisfies \texttt{isComplete}.

\camchg{However, in particular, \texttt{isComplete} can be various, defined by inner classes, and has side effects. For example, \Cref{fig:java-mutant-nameless-accumulate-test} shows the \texttt{Force} class implementation used in the test case \texttt{applyTimed}.}

For the \texttt{Torque} class, the situation is similar.

Therefore, killing the mutant that disturbs the \texttt{isComplete}-related feature is non-trivial. It cannot be called directly. Without calling it, validating is also non-trivial due to the various implementations with inner classes involved.

\clearpage

\begin{codefigure}{Python iterator-related filtered method example: \texttt{prefix\_lines} from miyuchina/mistletoe.}{fig:python-filter-iterator-prefix-lines}
\begin{Verbatim}[breaklines, breakanywhere]
@classmethod
def prefix_lines(
    cls,
    lines: Iterable[str],
    first_line_prefix: str,
    following_line_prefix: str = None,
) -> Iterable[str]:
    ...
    for line in lines:
        ...
        yield prefixed if not prefixed.isspace() else ""
\end{Verbatim}
\end{codefigure}

\begin{codefigure}{Python timeout-related filtered method example: \texttt{load\_inventory} from hynek/doc2dash.}{fig:python-filter-timeout-load-inventory}
\begin{Verbatim}[breaklines, breakanywhere]
def load_inventory(source: Path) -> Mapping[str, Mapping[str, InventoryEntry]]:
    ...
    with (source / "objects.inv").open("rb") as fp:
        ...
\end{Verbatim}
\end{codefigure}

\begin{codefigure}{Python concurrency-related filtered method example: \texttt{execute\_code} from vndee/llm-sandbox.}{fig:python-filter-concurrency-execute-code}
\begin{Verbatim}[breaklines, breakanywhere]
def execute_code(
    code: str,
    language: str = "python",
    libraries: list[str] | None = None,
    timeout: int = 30,
) -> list[ImageContent | TextContent]:
    ...
    try:
        ...
        with session_cls(
            ...
        ) as session:
            result = session.run(
                code=code,
                libraries=libraries or [],
                timeout=timeout,
            )
    ...
    else:
        return results
\end{Verbatim}
\end{codefigure}

\begin{codefigure}{Python coroutine-related filtered method example: \texttt{get\_task} from a2aproject/a2a-python.}{fig:python-filter-concurrency-get-task}
\begin{Verbatim}[breaklines, breakanywhere]
async def get_task(self) -> Task | None:
    ...
    self._current_task = await self.task_store.get(
        self.task_id, self._call_context
    )
    ...
    return self._current_task
\end{Verbatim}
\end{codefigure}

\begin{codefigure}{Python language-entity filtered method example: \texttt{while\_until\_true} from pypyr/pypyr.}{fig:python-filter-language-while-until-true}
\begin{Verbatim}[breaklines, breakanywhere]
def while_until_true(interval, max_attempts):
    def decorator(f):
        ...

    return decorator
\end{Verbatim}
\end{codefigure}

\begin{codefigure}{Python language-entity filtered method example: \texttt{make\_sync} from run-llama/llama\_deploy.}{fig:python-filter-language-make-sync}
\begin{Verbatim}[breaklines, breakanywhere]
def make_sync(_class: type[T]) -> Any:
    """Wraps the methods of the given model class so that they can be called without `await`."""

    class ModelWrapper(_class):  # type: ignore
        _instance_is_sync: bool = True
    ...
    return ModelWrapper
\end{Verbatim}
\end{codefigure}

\begin{codefigure}{Python ordinary-category filtered method example: \texttt{rich\_format\_error} from fastapi/typer.}{fig:python-filter-ordinary-rich-format-error}
\begin{Verbatim}[breaklines, breakanywhere]
def rich_format_error(self: click.ClickException) -> None:
    ...
    if ctx is not None:
        console.print(ctx.get_usage())

    if ctx is not None and ctx.command.get_help_option(ctx) is not None:
        console.print(
            ...
        )

    console.print(
        ...
    )
\end{Verbatim}
\end{codefigure}

\begin{codefigure}{Python specified-error-process example: test case covering \texttt{\_make\_command\_help} from langchain-ai/langchain-mcp-adapters.}{fig:python-filter-specified-error-help-create}
\begin{Verbatim}[breaklines, breakanywhere]
def test_help_create():
    result = runner.invoke(app, ["create", "--help"])
    assert result.exit_code == 0
    ...
\end{Verbatim}
\end{codefigure}

\begin{codefigure}{Python other filtered method example: \texttt{merge} from pypyr/pypyr.}{fig:python-filter-others-merge}
\begin{Verbatim}[breaklines, breakanywhere]
def merge(self, add_me):
    """Merge add_me into context and applies interpolation.
    ... (27 lines of comments)
    """
    def merge_recurse(current, add_me):
        ... (47 lines)
    ... (2 lines)
    merge_recurse(self, add_me)
\end{Verbatim}
\end{codefigure}

\begin{codefigure}{Python mutant-level ordinary example: \texttt{getprops} from google/mobly.}{fig:python-mutant-ordinary-getprops}
\begin{Verbatim}[breaklines, breakanywhere]
def getprops(self, prop_names):
    ...
    if attempt < attempts - 1:
        time.sleep(DEFAULT_GETPROPS_RETRY_SLEEP_SEC)
    return results
\end{Verbatim}
\end{codefigure}

\begin{codefigure}{Python mutant-level specified-error-process example: triggered test case for \texttt{\_parse\_ignore\_file} from coderamp-labs/gitingest.}{fig:python-mutant-specified-error-parse-ignore}
\begin{Verbatim}[breaklines, breakanywhere]
result = _invoke_isolated_cli_runner(["./", "--output", "-", "--exclude-pattern", "tests/"])
...
assert result.exit_code == 0
\end{Verbatim}
\end{codefigure}

\begin{figure*}[p]
\centering
\begin{subfigure}{\textwidth}
\centering
\begin{tcolorbox}[width=0.96\textwidth,colback=white,colframe=black,boxrule=0.5pt,arc=0pt,left=6pt,right=6pt,top=6pt,bottom=6pt]
\footnotesize
\begin{Verbatim}[breaklines, breakanywhere]
def k_closest(points, k, origin=(0, 0)):
    ...
    heapify(heap)
    ...
    for point in points[k:]:
        ...
    return [point for nd, point in heap]
\end{Verbatim}
\end{tcolorbox}
\caption{Original method.}
\end{subfigure}

\vspace{0.75em}

\begin{subfigure}{\textwidth}
\centering
\begin{tcolorbox}[width=0.96\textwidth,colback=white,colframe=black,boxrule=0.5pt,arc=0pt,left=6pt,right=6pt,top=6pt,bottom=6pt]
\footnotesize
\begin{Verbatim}[breaklines, breakanywhere]
def k_closest(points, k, origin=(0, 0)):
    ...
    heap.sort()
    ...
    for point in points[k:]:
        ...
    return [point for nd, point in heap]
\end{Verbatim}
\end{tcolorbox}
\caption{Equivalent mutant.}
\end{subfigure}

\vspace{0.75em}

\begin{subfigure}{\textwidth}
\centering
\begin{tcolorbox}[width=0.96\textwidth,colback=white,colframe=black,boxrule=0.5pt,arc=0pt,left=6pt,right=6pt,top=6pt,bottom=6pt]
\footnotesize
\begin{Verbatim}[breaklines, breakanywhere]
self.assertEqual(
    [(-2, -2), (1, 1), (1, 0), (-1, 0)], k_closest(...)
)
\end{Verbatim}
\end{tcolorbox}
\caption{Over-constrained test case.}
\end{subfigure}
\caption{\camcaption{Python mutant-level equivalent-result example: original method, equivalent mutant, and test case for \texttt{k\_closest} from keon/algorithms.}}
\label{fig:python-mutant-equivalent-k-closest}
\end{figure*}

\begin{figure*}[p]
\centering
\begin{subfigure}{\textwidth}
\centering
\begin{tcolorbox}[width=0.96\textwidth,colback=white,colframe=black,boxrule=0.5pt,arc=0pt,left=6pt,right=6pt,top=6pt,bottom=6pt]
\footnotesize
\begin{Verbatim}[breaklines, breakanywhere]
def partition_by_spaces(text, spaces) -> tuple[str, str, str]:
    ...
    for i, c in enumerate(text):
        ...
        elif start >= 0:
            end = i
            break
    ...
\end{Verbatim}
\end{tcolorbox}
\caption{Target method.}
\end{subfigure}

\vspace{0.75em}

\begin{subfigure}{\textwidth}
\centering
\begin{tcolorbox}[width=0.96\textwidth,colback=white,colframe=black,boxrule=0.5pt,arc=0pt,left=6pt,right=6pt,top=6pt,bottom=6pt]
\footnotesize
\begin{Verbatim}[breaklines, breakanywhere]
def parse(...) -> list[str] | SetextHeading:
    ...
    while ...:
        if cls.break_paragraph(source):
            break
        ...

def break_paragraph(...) -> bool:
    try:
        str1, str2, str3 = parse_leading(...)
        if ...:
            return True
        return False
    finally:
        ...
\end{Verbatim}
\end{tcolorbox}
\caption{Simplified downstream call path.}
\end{subfigure}
\caption{\camcaption{Python mutant-level timeout example: \texttt{partition\_by\_spaces} from frostming/marko and a simplified downstream call path.}}
\label{fig:python-mutant-timeout-partition-by-spaces}
\end{figure*}

\begin{codefigure}{Java timeout-related filtered method example: \texttt{compare} from red6/pdfcompare.}{fig:java-filter-timeout-compare}
\begin{Verbatim}[breaklines, breakanywhere]
public T compare() throws IOException, RenderingException {
    ...
    try (PDDocument expectedDocument = Loader.loadPDF(...)) {
        try (PDDocument actualDocument = Loader.loadPDF(...)) {
            compare(expectedDocument, actualDocument);
        }
    }
    ...
    return compareResult;
}
\end{Verbatim}
\end{codefigure}

\begin{codefigure}{Java iterator-related filtered method example: \texttt{iterator} from decorators-squad/eo-yaml.}{fig:java-filter-iterator-eo-yaml}
\begin{Verbatim}[breaklines, breakanywhere]
public Iterator<YamlLine> iterator() {
    Iterator<YamlLine> iterator = this.yamlLines.iterator();
    if (iterator.hasNext()) {
        final List<YamlLine> docsStart = ...
        ...
        iterator = docsStart.iterator();
    }
    return iterator;
}
\end{Verbatim}
\end{codefigure}

\begin{codefigure}{Java inaccessible-code-context filtered method example: \texttt{drawRectanglesOfDifferences} from romankh3/image-comparison.}{fig:java-filter-inaccessible-draw-rectangles}
\begin{Verbatim}[breaklines, breakanywhere]
private void drawRectanglesOfDifferences(List<Rectangle> rectangles, Graphics2D graphics) {
    List<Rectangle> rectanglesForDraw;
    ...
    draw(graphics, rectanglesForDraw);
    if (fillDifferenceRectangles) {
        fillRectangles(graphics, ...);
    }
}
\end{Verbatim}
\end{codefigure}

\begin{figure*}[p]
\centering
\begin{subfigure}{\textwidth}
\centering
\begin{tcolorbox}[width=0.96\textwidth,colback=white,colframe=black,boxrule=0.5pt,arc=0pt,left=6pt,right=6pt,top=6pt,bottom=6pt]
\footnotesize
\begin{Verbatim}[breaklines, breakanywhere]
public void changeMess() {
    for (int y : message) {
        double x = randomGenerator.nextDouble();
        ...
        if (x < ber) {
            messageChanged = true;
            ...
            message.set(...);
        }
    }
    if (messageChanged) {
        wrongMess++;
    }
}
\end{Verbatim}
\end{tcolorbox}
\caption{Target method.}
\label{fig:java-filter-random-change-mess}
\end{subfigure}

\vspace{0.75em}

\begin{subfigure}{\textwidth}
\centering
\begin{tcolorbox}[width=0.96\textwidth,colback=white,colframe=black,boxrule=0.5pt,arc=0pt,left=6pt,right=6pt,top=6pt,bottom=6pt]
\footnotesize
\begin{Verbatim}[breaklines, breakanywhere]
@Test
void testIntermediateBER() {
    CRCAlgorithm c = new CRCAlgorithm("1101", 4, 0.5);
    c.generateRandomMess();
    for (int i = 0; i < 1000; i++) {
        ...
        c.changeMess();
        ...
    }
    assertTrue(c.getWrongMess() > 0);
    ...
}
\end{Verbatim}
\end{tcolorbox}
\caption{Representative test case.}
\label{fig:java-filter-random-change-mess-test}
\end{subfigure}
\caption{\camcaption{Java random-related filtered method examples for \texttt{changeMess} from TheAlgorithms/Java.}}
\end{figure*}

\begin{codefigure}{Java ordinary-category filtered method example: \texttt{execute} from antkorwin/xsync.}{fig:java-filter-ordinary-execute}
\begin{Verbatim}[breaklines, breakanywhere]
public void execute(KeyT firstKey, KeyT secondKey, Runnable runnable) {
    ...
    if (...) {
        synchronized (firstMutex) {
            synchronized (secondMutex) {
                runnable.run();
            }
        }
    } else {
        synchronized (globalLock) {
            synchronized (firstMutex) {
                synchronized (secondMutex) {
                    runnable.run();
                }
            }
        }
    }
}
\end{Verbatim}
\end{codefigure}

\begin{codefigure}{Java other filtered method example: \texttt{douglasPeucker} from dyn4j/dyn4j.}{fig:java-filter-others-douglas-peucker}
\begin{Verbatim}[breaklines, breakanywhere]
/**
 * Recursively ... (8 lines of comments)
 */
private final List<Vector2> douglasPeucker(...) {
    ... (22 lines)
    if (...) {
        // sub-divide and run the algo on each half
        List<Vector2> aReduced = this.douglasPeucker(...);
        List<Vector2> bReduced = this.douglasPeucker(...);
        ... (4 lines)
    } else {
        ... (41 lines)
    }
    
    return result;
}
\end{Verbatim}
\end{codefigure}

\begin{figure*}[p]
\centering
\begin{subfigure}{\textwidth}
\centering
\begin{tcolorbox}[width=0.96\textwidth,colback=white,colframe=black,boxrule=0.5pt,arc=0pt,left=6pt,right=6pt,top=6pt,bottom=6pt]
\footnotesize
\begin{Verbatim}[breaklines, breakanywhere]
protected void accumulate(double elapsedTime) {
    ...
    if (...) {
        Iterator<Force> it = this.forces.iterator();
        while(it.hasNext()) {
            Force force = it.next();
            ...
            if (force.isComplete(elapsedTime)) {
                it.remove();
            }
        }
    }
    ...
    if (...) {
        Iterator<Torque> it = this.torques.iterator();
        while(it.hasNext()) {
            Torque torque = it.next();
            ...
            if (torque.isComplete(elapsedTime)) {
                it.remove();
            }
        }
    }
}
\end{Verbatim}
\end{tcolorbox}
\caption{Target method.}
\label{fig:java-mutant-nameless-accumulate}
\end{subfigure}

\vspace{0.75em}

\begin{subfigure}{\textwidth}
\centering
\begin{tcolorbox}[width=0.96\textwidth,colback=white,colframe=black,boxrule=0.5pt,arc=0pt,left=6pt,right=6pt,top=6pt,bottom=6pt]
\footnotesize
\begin{Verbatim}[breaklines, breakanywhere]
@Test
public void applyTimed() {
    ...
    Force f = new Force(1, 0) {
        private double time = 0;
        public boolean isComplete(double elapsedTime) {
            time += elapsedTime;
            if (time >= 2.0 / 60.0) {
                return true;
            }
            return false;
        }
    };
    ...
}
\end{Verbatim}
\end{tcolorbox}
\caption{Representative test implementation.}
\label{fig:java-mutant-nameless-accumulate-test}
\end{subfigure}
\caption{\camcaption{Java mutant-level nameless-implementation examples from dyn4j/dyn4j.}}
\end{figure*}

\clearpage

\begin{figure*}[htbp] % h=here t=top b=bottom p=page of floats
  \centering
  \includegraphics[width=0.6\linewidth]{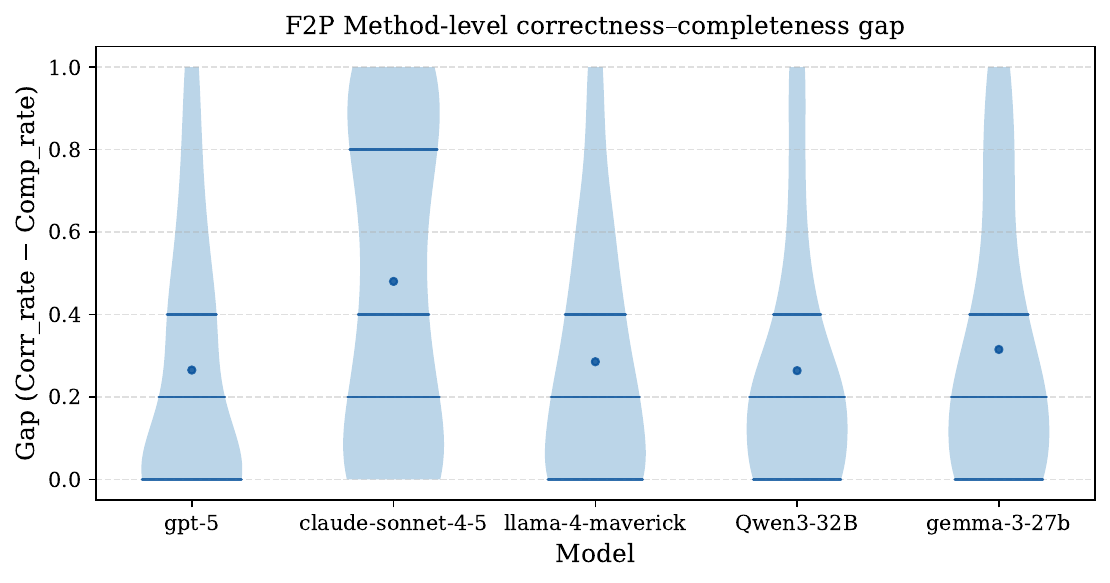}
  \caption{
  Distribution of method-level correctness–completeness gaps with F2P setting under different models.
  For each method $m$, the gap is computed as $\Delta(m) = \text{corr\_rate}(m) - \text{comp\_rate}(m)$, where corr\_rate and comp\_rate denote the fraction of generated postcondition sets that are test-correct and complete, respectively. 
  Each violin aggregates gap values across methods for which at least one test-correct postcondition is generated; methods with Corr@k = 0 are excluded.
  % Each violin aggregates gap values across all methods, pooling results from all models and both languages. 
  }
  \label{fig:violin--app-1}
\end{figure*}

\begin{figure*}[htbp] % h=here t=top b=bottom p=page of floats
  \centering
  \includegraphics[width=0.8\linewidth]{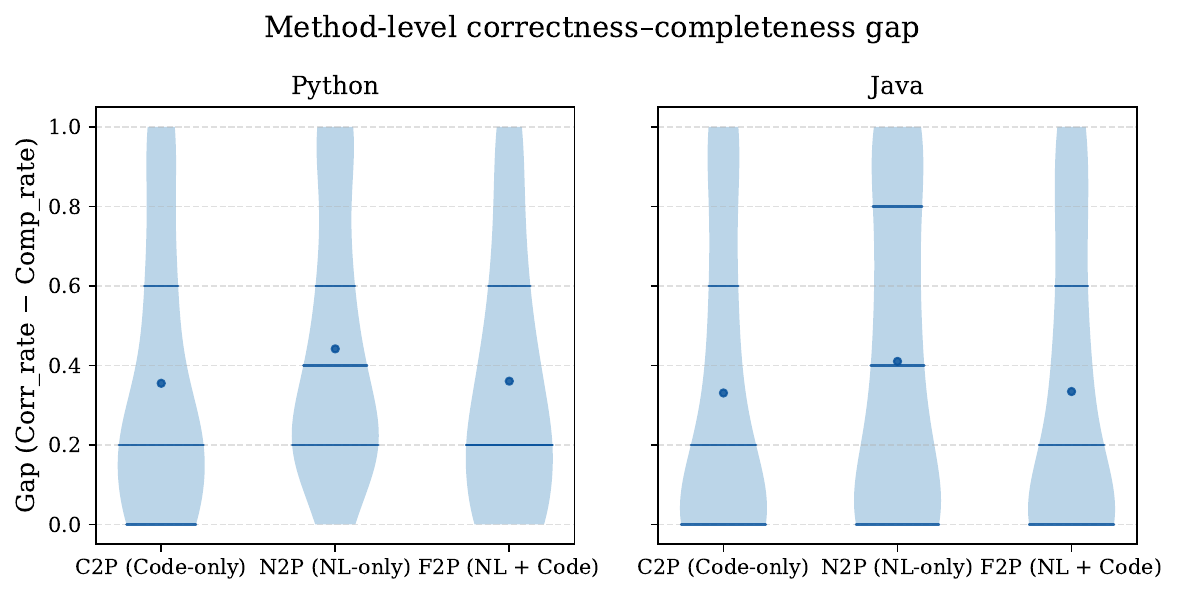}
  \caption{
  Distribution of method-level correctness–completeness gaps under different input settings and programming languages.
  For each method $m$, the gap is computed as $\Delta(m) = \text{corr\_rate}(m) - \text{comp\_rate}(m)$, where corr\_rate and comp\_rate denote the fraction of generated postcondition sets that are test-correct and complete, respectively. 
  Each violin aggregates gap values across methods for which at least one test-correct postcondition is generated; methods with Corr@k = 0 are excluded.
  % Each violin aggregates gap values across all methods, pooling results from all models and both languages. 
  }
  \label{fig:violin--app-2}
\end{figure*}

\end{document}